\documentclass[11pt,aps,prd,nofootinbib,superscriptaddress, onecolumn,preprintnumbers,balancelastpage]{revtex4}
\usepackage{amssymb,amsmath,latexsym,graphics, graphicx,epsfig,multirow,comment,hyperref,appendix, verbatim} 

\preprint{FERMILAB-PUB-11-311-T}

\newcommand{\lsim}{ \mathop{}_{\textstyle \sim}^{\textstyle <}}

\newcommand{\be}{\begin{eqnarray}}
\newcommand{\ee}{\end{eqnarray}}

\newcommand{\gev}{\rm \, GeV}

\newcommand{\kev}{\rm \, keV}

\def\be{\begin{equation}}
\def\ee{\end{equation}}
\def\bea{\begin{eqnarray}}
\def\eea{\end{eqnarray}}

\def\ltap{\ \raise.3ex\hbox{$<$\kern-.75em\lower1ex\hbox{$\sim$}}\ }
\def\gtap{\ \raise.3ex\hbox{$>$\kern-.75em\lower1ex\hbox{$\sim$}}\ }

\newcommand{\kmpers}{\mathrm{km/s}}
\newcommand{\elow}{\text{E}_{\text{low}}}
\newcommand{\ehigh}{\text{E}_{\text{high}}}

\begin{document}

\title{A CoGeNT Modulation Analysis}

\author{Patrick J. Fox$^a$, Joachim Kopp$^a$, Mariangela Lisanti$^b$, Neal Weiner$^{c,d}$\\
$^a$ Theoretical Physics Department, Fermi National Accelerator Laboratory, Batavia, Illinois, USA\\
$^b$ PCTS, Princeton University, Princeton, NJ 08540\\
$^c$ Center for Cosmology and Particle Physics, Department of Physics, New York University, NY, NY\\
$^d$ School of Natural Sciences, Institute for Advanced Study, Einstein Drive, Princeton, NJ 08540}

\begin{abstract}
We analyze the recently released CoGeNT data with a focus on their time-dependent properties.  Using a variety of statistical tests, we confirm the presence of modulation in the data, and find a significant component at high ($E_{ee} \gtap 1.5$ keVee) energies. We find that standard elastic WIMPs in a Maxwellian halo do not provide a good description of the modulation. We consider the possibility of non-standard halos, using halo independent techniques, and find a good agreement with the DAMA modulation for Q$_{\text{Na}} \approx 0.3$, but disfavoring interpretations with Q$_{\text{Na}}$ = 0.5. The same techniques indicate that CDMS-Ge should see an $O(1)$ modulation, and XENON100 should have seen 10-30 events (based upon the modulation in the 1.5-3.1 keVee range), unless L$_{\text{eff}}$ is smaller than recent measurements. Models such as inelastic dark matter provide a good fit to the modulation, but not the spectrum. We note that tensions with XENON could be alleviated in such models if the peak is dominantly in April, when XENON data are not available due to noise.
\end{abstract}

\maketitle
\newpage
\section{Introduction}
\label{sec:intro}

The CoGeNT Collaboration has recently published results from the first fifteen months of data taking~\cite{Aalseth:2010vx, Aalseth:2011wp}.  Since their first data release more than a year ago, they continue to observe an õunexplained excess in the spectrum of nuclear recoil scattering rate and now claim an annual modulation of 2.8$\sigma$\cite{Aalseth:2011wp}.  This preliminary evidence for a modulation is an important step towards determining the nature of CoGeNT's unexplained spectrum and has been claimed to be evidence for a $\sim7$ GeV dark matter~\cite{Hooper:2011hd}.  In this work, we present a comprehensive statistical analysis of the CoGeNT data and show that the modulation spectrum is hard to achieve with a conventional light elastic WIMP in a standard Maxwellian halo.    

Direct detection experiments such as CoGeNT search for the scattering of dark matter off nuclei in ground-based detectors.  The spectrum of nuclear recoil energies depends on the mass and scattering cross section of the dark matter, as well as its velocity distribution in the Galaxy.  One of the most distinctive features of a dark matter signal is that it should modulate annually due to the motion of the Earth's rotation about the Sun~\cite{Drukier:1986tm}.  In particular, the flux of dark matter as observed in the lab frame is larger in the summer, when the Earth is moving in the same direction as the Sun, than in the winter, when the Earth's motion is against that of the Sun~\cite{Lewin:1995rx}.  For a Maxwell-Boltzmann velocity distribution, the flux peaks 152 days into the year. 

Observing an annual modulation in a potential signal is a crucial step in confirming its origin as dark matter.  Direct detection experiments face the challenge of distinguishing dark matter nuclear recoils from a list of potential backgrounds.  In most cases, experiments utilize a combination of ionization, scintillation, or phonon signals to separate out nuclear recoils~\cite{Gaitskell:2004gd}.  But the possibility of contamination in the nuclear recoil band remains, for instance due to unaccounted for radioactive decays.  If the signal in the nuclear recoil band modulates with the period and phase expected for a dark matter signal, however, it would be a strong indication of the nature of the interaction producing the signal.     

To date, only the DAMA~\cite{Bernabei:2008yi,Bernabei:2010mq} and CoGeNT~\cite{Aalseth:2010vx, Aalseth:2011wp} experiments have claimed an annual modulation signal.  The DAMA experiment, which uses target crystals of NaI(Tl), claims an $8.9\sigma$ modulation with period $0.999\pm0.002$ years and peaking at $146\pm7$ days.  The CoGeNT experiment, which uses a Ge target, has recently claimed an annual modulation signal with their first fifteen months of data.  For energies ranging from 0.5-3.0~keVee,\footnote{The notation keVee refers to the ``electron equivalent energy in keV'', which is defined as the reconstructed recoil energy under the assumption that it is carried by an electron. For nuclear recoils, only part of the recoil energy is visible in the detector---an effect that has to be corrected for by dividing the visible energy by a quenching factor---so that the energy threshold for nuclear recoils is higher than that for electron recoils. When referring to a nuclear recoil energy, we will use the notation ``keVnr''.} they observe a maximal modulation with best-fit modulation fraction of $16.6\pm3.8\%$, period $347\pm29$ days, and minimum at Oct. 16$\pm 12$ days.  In this energy range, the significance is $2.8\sigma$.  When the CoGeNT data are fit with a dark matter signal in addition to a constant and exponential background, the best-fit dark matter mass is roughly 7-8 GeV with $\sigma\sim 10^{-4}$ pb, which is close to the region of parameter space that is consistent with DAMA~\cite{Aalseth:2010vx, Hooper:2011hd, Arina:2011si}.

The light dark matter interpretation of CoGeNT and DAMA has been challenged by null results from other direct detection experiments, such as XENON100~\cite{Aprile:2011hi}, XENON10~\cite{Angle:2007uj,Angle:2011th}, Simple~\cite{Felizardo:2010mi,Felizardo:2011uw}, and CDMS~\cite{Ahmed:2009zw,Ahmed:2010wy}.  The compatibility of the XENON and CoGeNT results has been discussed in greater detail in~\cite{Collar:2011wq,Collar:2010ht,Collar:2010nx,Collar:2010gd,Collar:2010gg}.  Reconciling CDMS and CoGeNT is more challenging because the two use the same target material and CDMS has reported an event rate significantly below that of CoGeNT in the same energy range; however, it has been claimed that errors in the energy calibration can potentially cause the discrepancy~\cite{Collar:2011kf}. As we shall see, our conclusions on the modulation at CoGeNT will not depend strongly on these details.   

This work presents a detailed statistical analysis of the CoGeNT results using the publicly available data~\cite{juanprivate}.
Section~\ref{sec:techniques} introduces the statistical tests that will be used throughout the paper.  Section~\ref{sec:modulation} presents a model-independent analysis of the modulated and unmodulated rate, period, and phase.  
Section~\ref{sec:darkmatter} discusses the implications of these results for dark matter  and shows that the hypothesis of a light, elastically-scattering WIMP (Weakly Interacting Massive Particle) is strained. In addition, the consistency of CoGeNT with CDMS-Ge, CDMS-Si, XENON100 and DAMA is presented using an analysis that is independent of astrophysical uncertainties.

\section{CoGeNT Data and Analysis Techniques}
\label{sec:techniques}

We perform our analysis on the data that is available from the CoGeNT Collaboration upon request~\cite{juanprivate}. The first event was recorded on December 4, 2009 and the full data run spanned 458~days, of which 442 were live.  The known background in the energy region of interest arises from cosmogenic L-shell electron capture events.  A complete description of the backgrounds, efficiencies, and experimental deadtime is included in Appendix~\ref{sec:gapsandbacks}.

A binned chi-squared analysis of the data is done to confirm the 2.8$\sigma$ significance for modulation in the 0.5-3.0 keVee energy bin~\cite{Aalseth:2011wp, Hooper:2011hd}.  In addition, we use two alternate statistical tests that are appropriate for searches of periodicities in data: the unbinned maximum likelihood method~\cite{Cowan:2008zza} and the Lomb-Scargle periodogram~\cite{Scargle:1982, Lomb:1976}.  Both these tests have been used, for example, in searches for periodicities in the solar neutrino flux from the Sudbury Neutrino Observatory~\cite{Aharmim:2005iu}.  Below, we review these statistical tests and note that, where the different techniques can be compared, they give qualitatively similar results.

\subsection{Binned Analysis}

We carry out a simple chi-squared analysis on events in a given energy range, binned in time.  Each event, located at energy $E_i$, is reweighted by the efficiency at that energy, $f_{\text{eff}}(E_i)^{-1}$.  The cosmogenic background contribution is determined  for each bin and then subtracted from the number of events in the bin.  A correction is applied to the time bins that overlap with the shutdown periods of the detector. The errors in each bin are treated as Gaussian,\footnote{This is a reasonable approximation for the number of time and energy bins used in this paper.} based on the original bin contents before reweighting or subtraction.  

The subtracted binned data are fit with a modulated spectrum of the form 
\be
R(t) = A_0 (1 + A_1 \cos(\omega(t - t_0 ))~,
\label{eq:modrate}
\ee 
where $\omega$ is the oscillation period,  $t_0$ is the phase, $A_1$ is the modulation fraction, and $A_0$ is the unmodulated rate.  Note that the unmodulated rate may contain contributions from a dark matter component, as well as any other constant backgrounds.  All times are taken relative to January 1$^{\mathrm{st}}$, 2010.  We consider several types of fits, including: (1) all parameters are allowed to float, (2) the period is fixed to one year, $\omega_0 = 2\pi/$year, (3) the period is fixed to $\omega_0$ and the phase is fixed to $t_0 = 152$ days, the value expected for dark matter in the standard halo model (SHM), and (4) the modulation fraction is set to zero (null hypothesis).

\subsection{Unbinned Analysis}

To maintain access to all the information in the time distribution of events, we carry out an unbinned maximum likelihood analysis.  This method can be used to test the hypothesis that the excess data
above backgrounds follow a rate distribution of the form in Eq.~\ref{eq:modrate}.  If no assumption is made about the energy distribution of the data, as in Sect.~\ref{sec:modulation}, then it should be binned in energy, but not in time.  The probability density function (PDF) for events in the energy range $(\text{E}_{\text{low}}, \text{E}_{\text{high}}=\text{E}_{\text{low}}+\Delta \text{E})$ is
\bea
\phi(t) & = & \left[ 0.33\,\mathrm{kg}\times \Delta\text{E}\, \bar{f}_{\text{eff}}(\text{E}) R(t) + \int_{\text{E}_{\text{low}}}^{\text{E}_{\text{high}}} f_{\text{cosmo}}(E, t)f_{\text{eff}}(\text{E}) \right] f_{\text{gaps}}(t)~,
\eea
where $\bar{f}_{\text{eff}}(\text{E})$ is the weighted average efficiency in the energy bin, $f_{\text{cosmo}}(E,t)$ is the model of the cosmogenic backgrounds, and $f_{\text{gaps}}(t)$ accounts for experimental deadtimes.  See Appendix~\ref{sec:gapsandbacks} for a more complete discussion of the background modeling.  In Sect.~\ref{sec:darkmatter}, we carry out fully binned and unbinned analyses to establish whether the data is explained by dark matter.   In this case, the PDF does not include the integral over energy.

Using this PDF, the extended (log-)likelihood is 
\begin{equation}
  2 \log L(A_0, A_1, \omega, t_0) = 2 \sum_{i} \phi(t_i) - 2 \int_{t_{\text{start}}}^{t_{\text{end}}}dt\, \phi(t)~,
\end{equation}
where $t_{\text{start}} = -28$~days and $t_{\text{end}}= 429.9$~days for the data run in~\cite{Aalseth:2011wp}, and the sum is over all data points in the sample.  The quantity $L(A_0, A_1, \omega, t_0)$ is maximized, subject to the constraints that $A_{0,1}$ be positive and that $0 \leq t_0\leq \mathrm{year}$.  The significance for any particular hypothesis relative to any other is the difference between $2 \log L_{\text{max}}$ for each; this difference follows a $\chi^2$ distribution.

\subsubsection{Lomb-Scargle Periodogram}

The weighted Lomb-Scargle technique is ideally suited to search for periodic signals in unevenly sampled data, such as that of CoGeNT.  For data that are divided into $N$ independent time bins with $y(t_i)$ data points each ($i=1,\ldots, N$), the Lomb-Scargle power for frequency $f$ is given by 
\begin{equation}
P(f) = \frac{1}{2\sigma^2} \Bigg( 
\frac{\big[ \sum_{i=1}^N W_i \, (y(t_i) - \bar{y}) \, \cos\,\omega (t_i - \tau) \big]^2}{\sum_{i=1}^N W_i \, \cos^2\,\omega (t_i - \tau)}+
 \frac{\big[ \sum_{i=1}^N W_i \, (y(t_i) - \bar{y}) \, \sin\,\omega (t_i - \tau) \big]^2}{\sum_{i=1}^N W_i \, \sin^2\,\omega (t_i - \tau)}
 \Bigg),
\end{equation}
where $\bar{y}$ and $\sigma$ are the weighted mean and variance for the data in all the time bins and $\omega$ is the angular frequency.  The phase factor $\tau$ and weight factor $W_i$ are given by 
\begin{equation}
\tan(2\omega \tau) = \frac{\sum_{i=1}^N W_i \, \sin\,2\omega t_i}{\sum_{i=1}^N W_i \, \cos\,2\omega t_i} \qquad \text{and} \qquad
W_i = \frac{1/\sigma_i^2}{\langle 1/\sigma_i^2 \rangle }~,
\end{equation}
respectively.  Here, $\sigma_i$ are the individual uncertainties in each bin.
\begin{figure}[t] 
   \centering
 \includegraphics[width=2.75in]{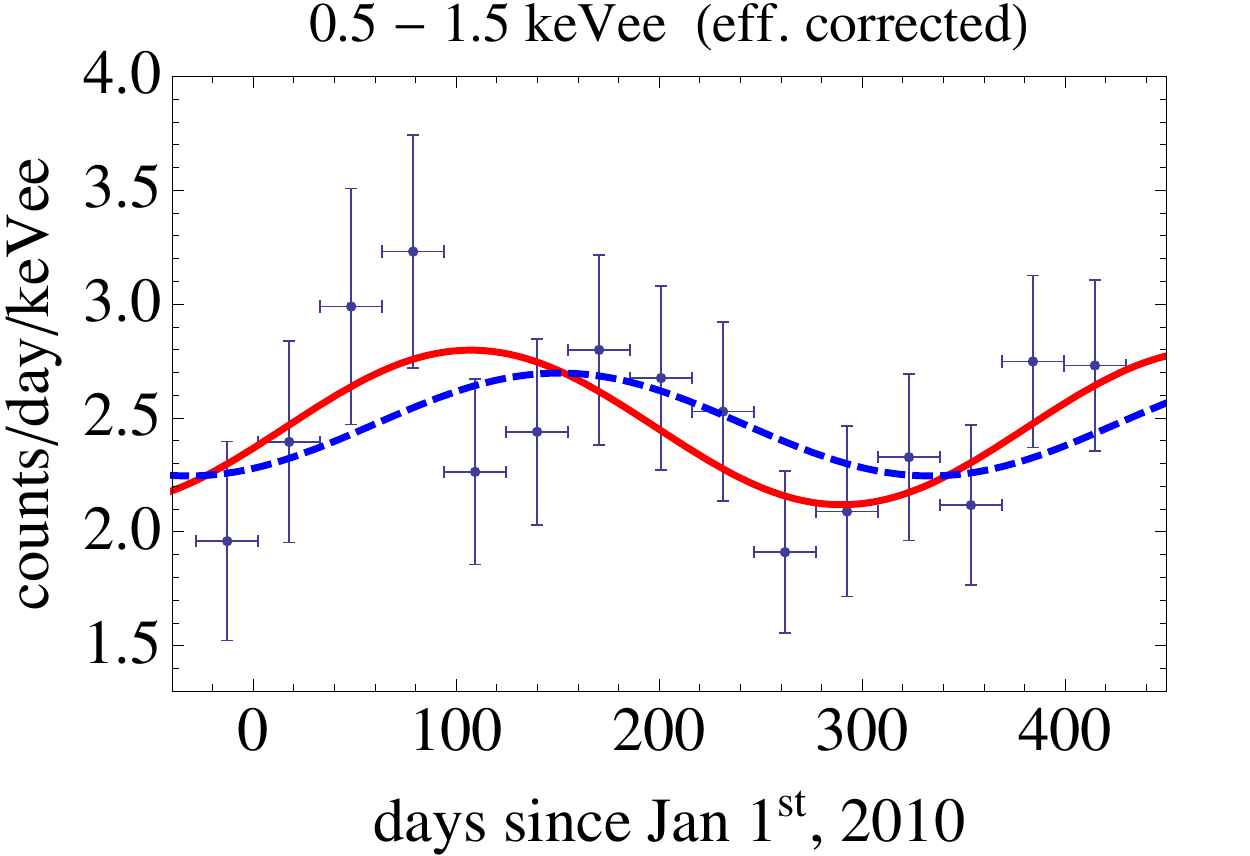}
  \includegraphics[width=2.75in]{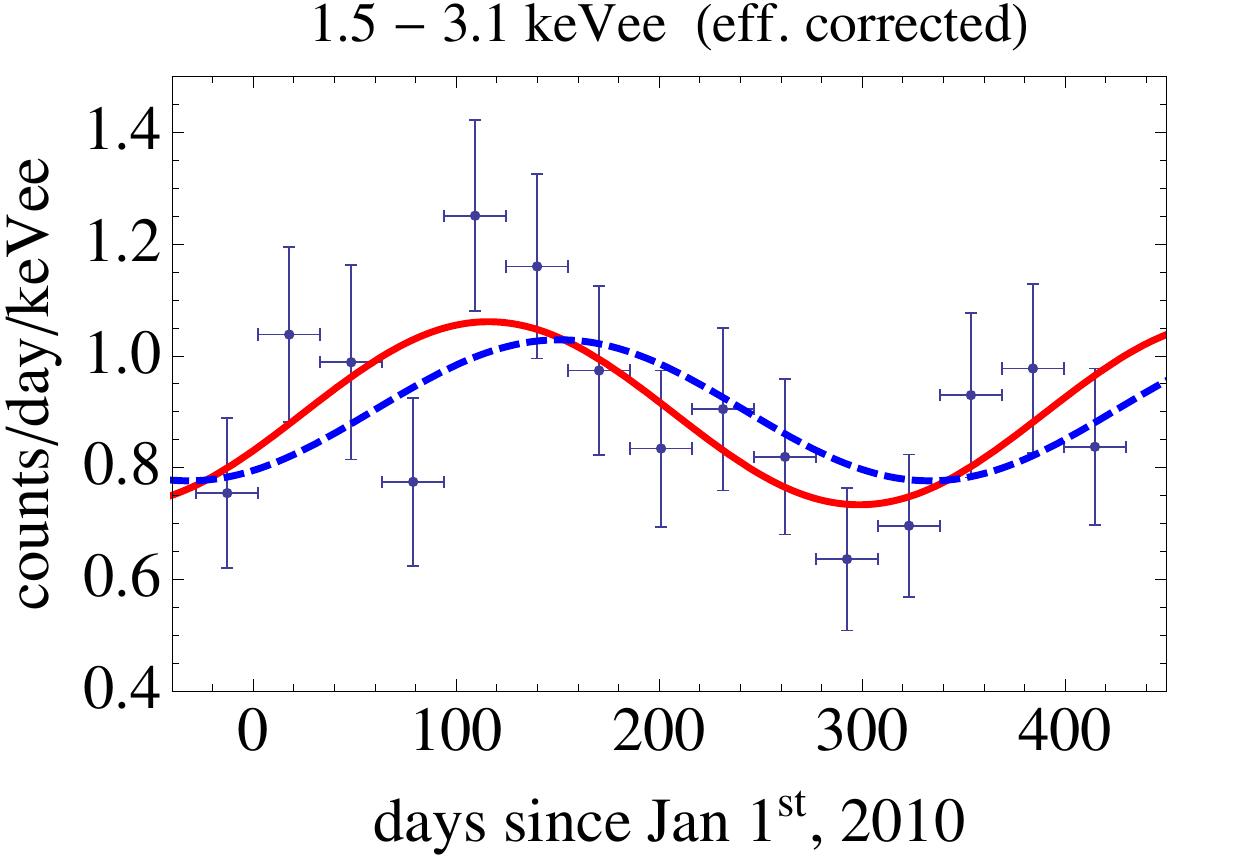}
   \caption{Time-binned data in various energy ranges. Specifically ({\it left}) [0.5--1.5]~keVee,  ({\it right}) [1.5--3.1]~keVee.  Overlaid are the best-fit to the modulation, as derived using the binned analysis, with free phase ({\it solid red curve}) and peak set at 152~days ({\it dashed blue}).  The best-fit points correspond to $A_0 = 7.4\ (7.5)$~events/day/kg/keVee, $A_1=0.14\ (0.09)$ and $t_0 = 107\ (152)$ days, for the phase free ($t_0=152$ days) for the lower bin and $A_0 = 2.7\ (2.7)$~events/day/kg/keVee, $A_1=0.18\ (0.14)$ and $t_0 = 116\ (152)$ days for the higher.}
   \label{fig:binning}
\end{figure}

For a given energy range, the events are divided into eighty time bins of approximately six days each.  In order for the Lomb-Scargle analysis to have a well-defined statistical interpretation, the contents of each bin must be large enough that the error on the number of events is well approximated by a Gaussian.  As a result, each bin is required to contain ten or more events.  A simple algorithm is used to merge any bin that contains fewer than ten events with the next highest bin.\footnote{If the last bin has fewer than ten events, it is merged with the penultimate bin.}  This procedure is repeated until no bin has fewer than ten events.  In addition, the centers of the bins are shifted to take into account any deadtime in the experiment.  Finally, the bin contents are efficiency-adjusted, the L-shell background in every bin is subtracted off, and the contents of the bin are converted to units of events/day/keVee.  The error is based on the total (pre-subtraction) bin contents.  This error is important for determining the weighting factors $W_i$.  The power observed in the frequency $\omega_0 = 2\pi /\text{year}$ can be converted to a significance for an oscillating signal.  The probability of observing power $P$ at any particular frequency in data that do not contain an oscillating signal is $e^{-P}$, whereas the probability for observing power $P$ at \emph{any} frequency (including the appropriate trial factor) is approximately $1 - (1 - e^{-P})^N$, where $N$ is the number of time bins~\cite{Press:NumRecip}.

\section{A Study of Modulation}
\label{sec:modulation}

The central goal of this work is to understand the properties of a potential modulation in the CoGeNT data.  Therefore, we begin by applying the statistical techniques presented above to analyze the properties of the modulation, without any assumptions of its origin.  We reproduce the results in~\cite{Aalseth:2011wp}, where a time-binned analysis  is done in the energy ranges 0.5--0.9~keVee and 0.5--3.0~keVee.  The results of~\cite{Aalseth:2011wp} suggest that the region above 0.9~keVee exhibits a sizeable component of the modulation.  For clarity, then, we divide the energy into two exclusive regions:  a ``low region'' [0.5--1.5 keVee] and a ``high region'' [1.5--3.1 keVee], shown in Figure~\ref{fig:binning}. 

\begin{figure}[b]
  \begin{center}
    \begin{tabular}{cc}
      \includegraphics[width=6cm]{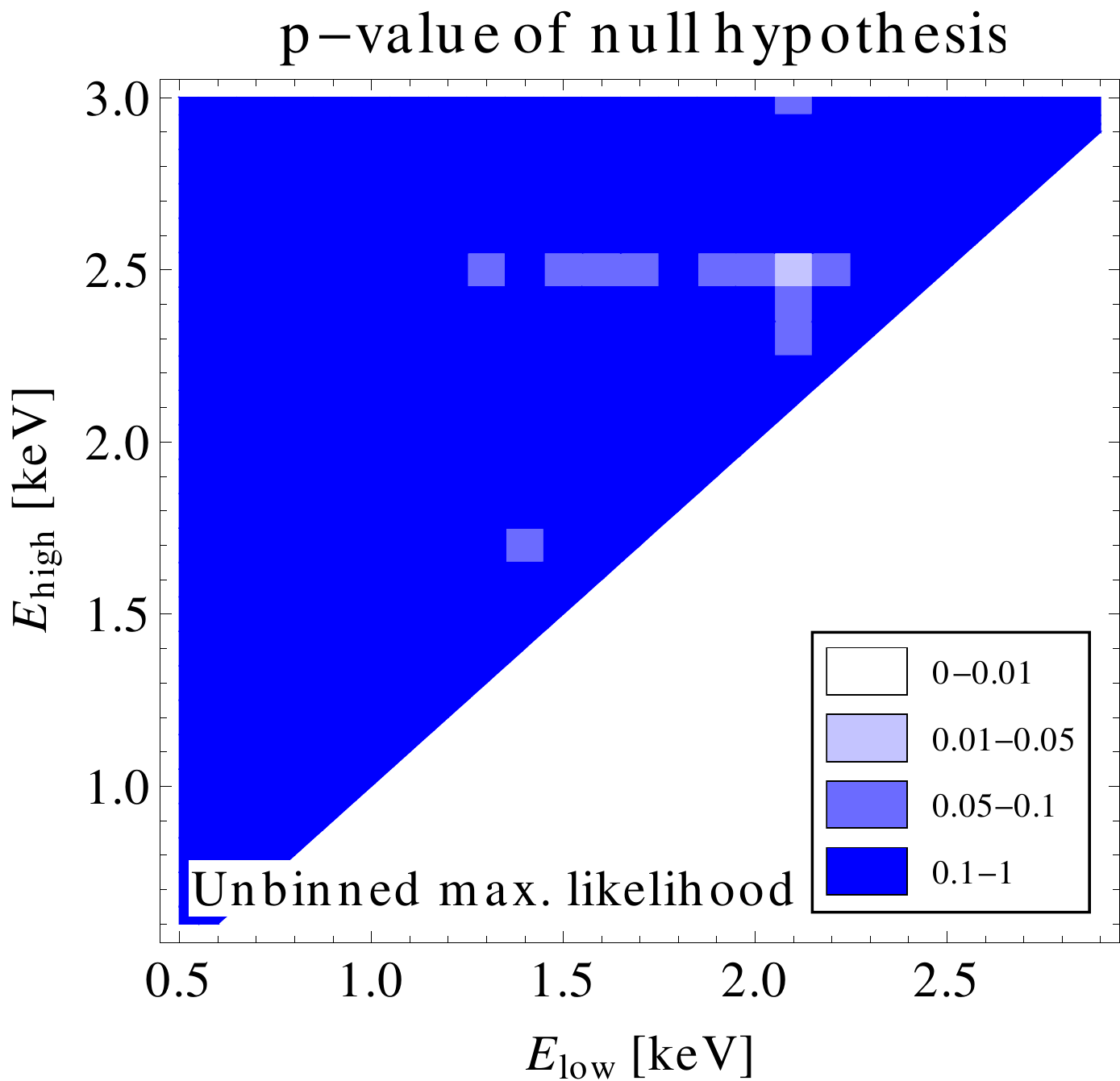} &
      \includegraphics[width=6cm]{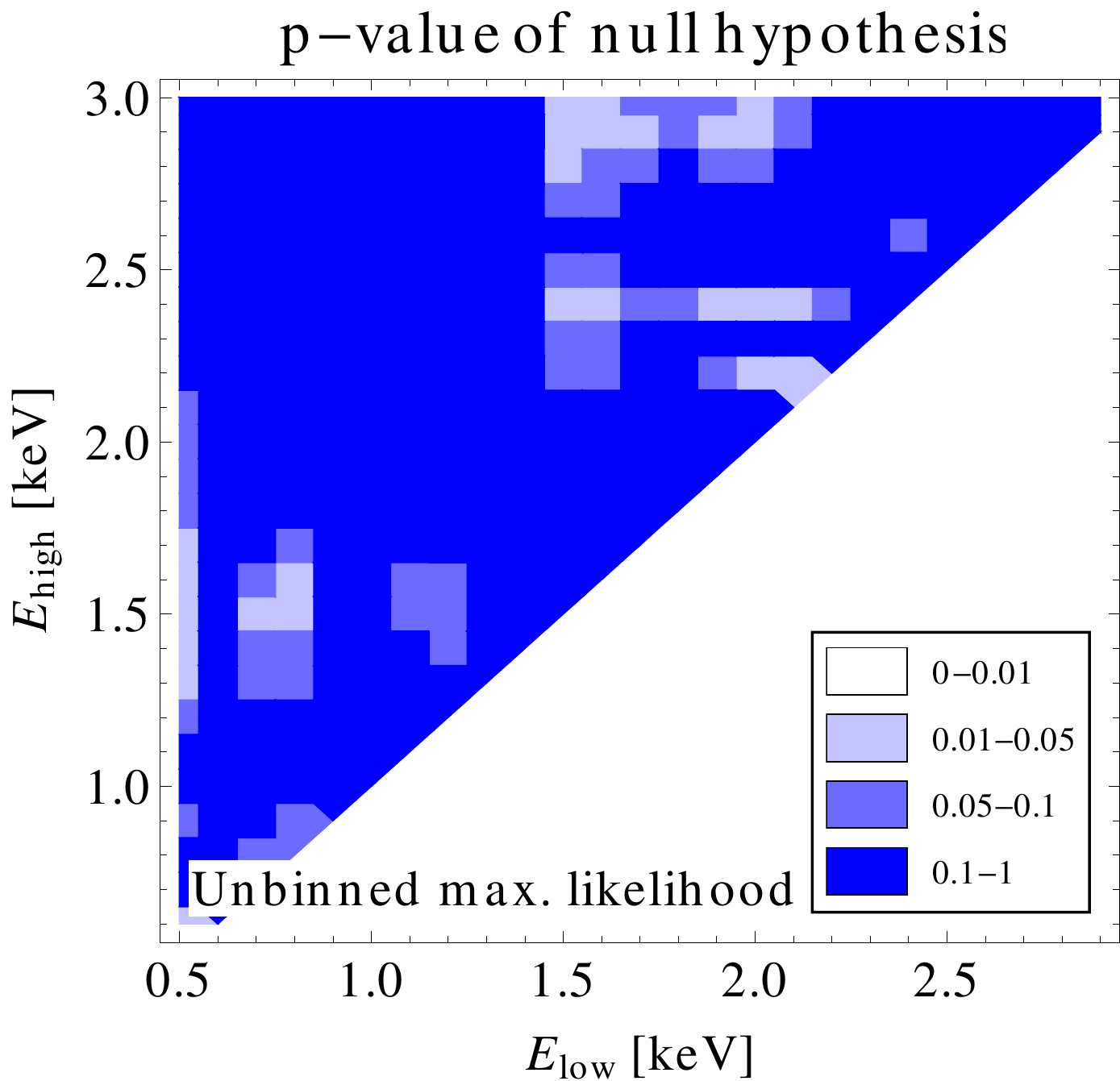} \\
      (a) & (b)
    \end{tabular}
  \end{center}
  \caption{Significance of daily modulation in CoGeNT, as measured by the probability for the null (no modulation) hypothesis to give the observed amount of modulation, in a bin with energies ranging from $\elow$ to $\ehigh$. We fit a model of the form \eqref{eq:modrate} to the data after subtracting backgrounds
    and correcting for detection efficiencies and shutdown periods.  The oscillation period $\omega$ is kept fixed at (a) one solar day (24~hrs) and
    (b) one sidereal day (23.93~hrs); the average rate $A_0$,
    the modulation fraction $A_1$, and the phase $t_0$ are free parameters in the fit.}
  \label{fig:daily}
\end{figure}

The data clearly exhibit a modulation over a wide range of energies.  The significance of adding a modulating term with a free phase (2 parameters), relative to the null hypothesis of no modulation, is $\Delta \chi^2 = $ 
4.7 in the range [0.5-1.5] keVee, and 
8.2 in the range [1.5-3.1] keVee.  Furthermore, the improvement in $\Delta \chi^2$ arising from adding a cosine with fixed phase (1 parameter) is 
2.3 for the [0.5-1.5] keVee range, and 
5.2 for [1.5-3.1] keVee. There is strong support for modulation in the high energy range, with little benefit in the low range.  In addition, the high energy region prefers a phase that is different from that expected for a Maxwellian halo.  

The modulation in energies above 1.5 keVee is surprising, as the rate spectrum in this region had previously been interpreted as a constant background contribution~\cite{Aalseth:2010vx}.  The unexpected nature of this modulation warrants a careful analysis of its properties and in the next two subsections, we apply additional tests to study its period, phase and amplitude.  The last subsection presents the energy spectrum for the unmodulated and modulated rates as well as for the phase, assuming an oscillation period of a year.  
\begin{figure}[b]
  \begin{center}
    \begin{tabular}{cc}
        \includegraphics[width=0.45\textwidth]{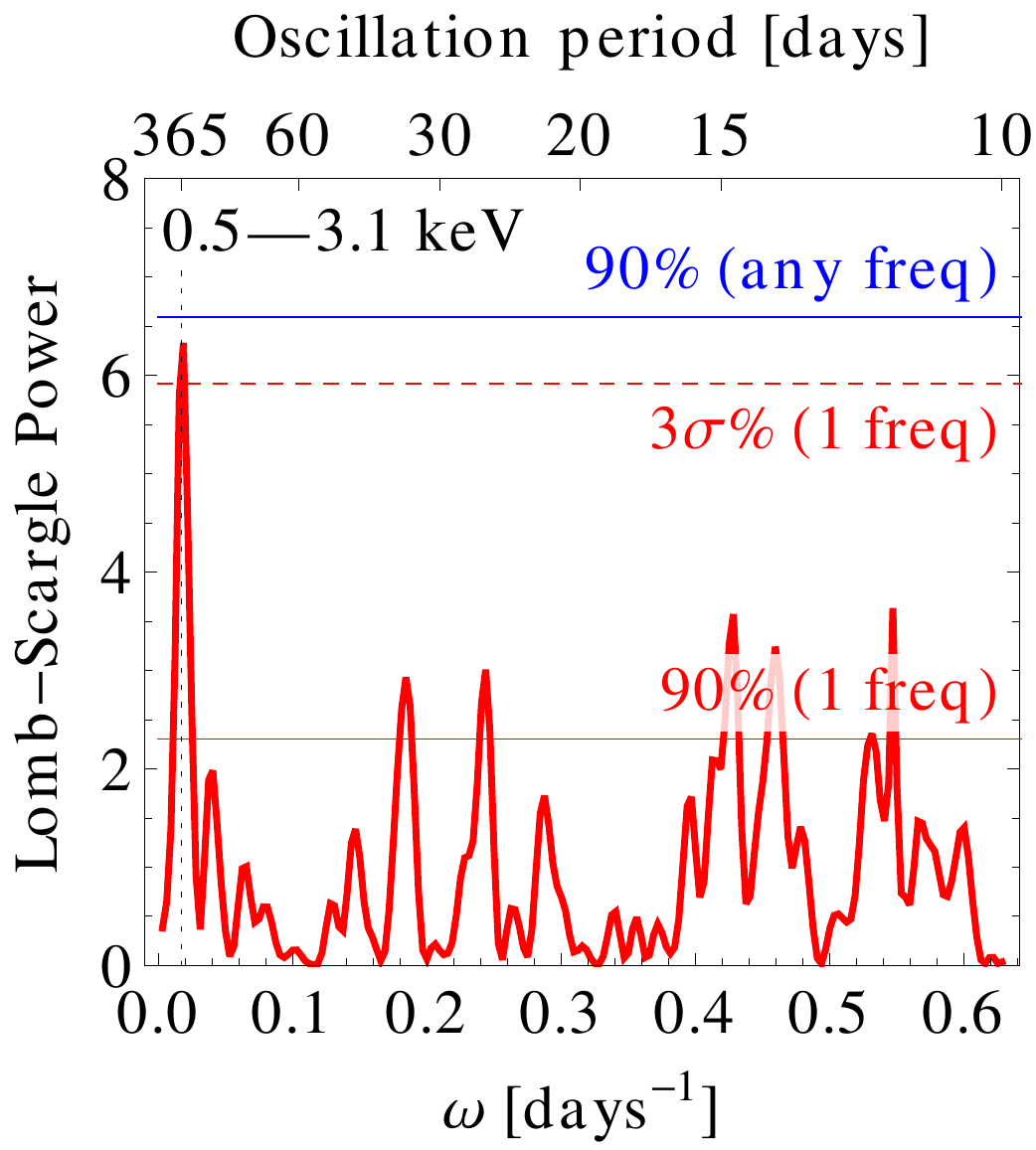} &
       \raisebox{0.4cm}{\includegraphics[width=0.44\textwidth]{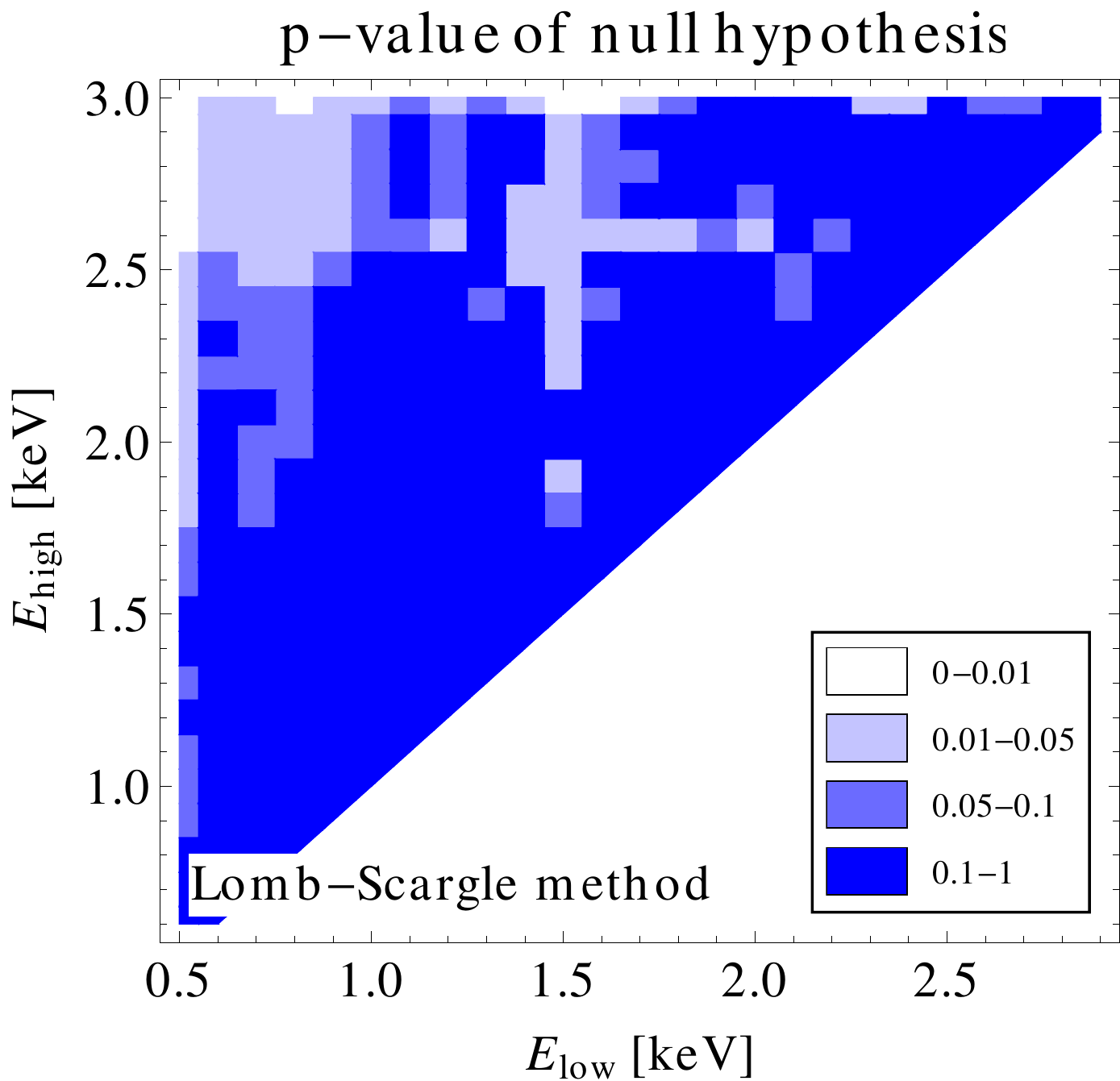}} \\
      (a) & (b)
    \end{tabular}
  \end{center}
  \caption{Results of the Lomb-Scargle analysis.  The left panel shows the Lomb-Scargle
    periodogram for the full energy range 0.5--3.1~keVee and the right panel
    plots the significance of annual modulation as a function of the considered energy range, from $\elow$ to $\ehigh$.}
  \label{fig:lomb-scargle}
\end{figure}

\subsection{Oscillation Period}

The first step in characterizing the CoGeNT modulation is to determine the relevant time periods that show up in the data.  The most obvious to check is evidence for daily modulation.  While the daily modulation expected from dark matter is negligible in a detector like CoGeNT, many sources of background, such as those induced by radon decays and cosmic rays, can depend on the time of day. For instance, at night the atmosphere is colder and denser so that the secondary pions produced in cosmic ray interactions are more likely to lose energy through scattering before they decay~\cite{Ambrosio:2002db}.

Figure~\ref{fig:daily} shows the significance of modulation, under
the assumption that the oscillation period is one solar day (24~hrs) and under the
assumption that it is one sidereal day (23.93~hrs). The plots show results for different energy ranges $E_{\text{low}}$ through $E_{\text{high}}$, where $E_{\text{low}}$ and $E_{\text{high}}$ range from 0.5 and 3.0~keVee in steps of 0.1~keVee. Even though the best-fit solutions
typically include about 10--20\% modulation, the statistical significance is
very small, as indicated by the $p$-values in the plots.  A few
isolated energy intervals exhibit a modulation with more
than $2\sigma$ confidence, but the significance is much lower once the trial factor for this to
happen anywhere in the considered energy range is included.  We thus conclude that the CoGeNT data do not show
evidence for diurnal modulation.

More generally, we can also search for modulation with any frequency using the Lomb-Scargle technique. The results are shown in Fig.~\ref{fig:lomb-scargle}.  The strongest modulation in the data has a period of one year; the Lomb-Scargle significance for annual modulation is above $3\sigma$ if no trial factor is included, and around 90\% with a trial factor.  The right panel of Fig.~\ref{fig:lomb-scargle} shows the significance of annual modulation, defined as the probability of obtaining the observed annual modulation from statistical fluctuations alone (not including a trial factor), as a function of the considered energy range from $E_{\text{low}}$ to $E_{\text{high}}$.  There is no significant modulation below $\sim 1.7$~keVee, but the significance increases once higher energies are included.

\subsection{Phase and Amplitude for Annual Modulation}
\begin{figure}[b] 
   \centering
   \includegraphics[width=2.0in]{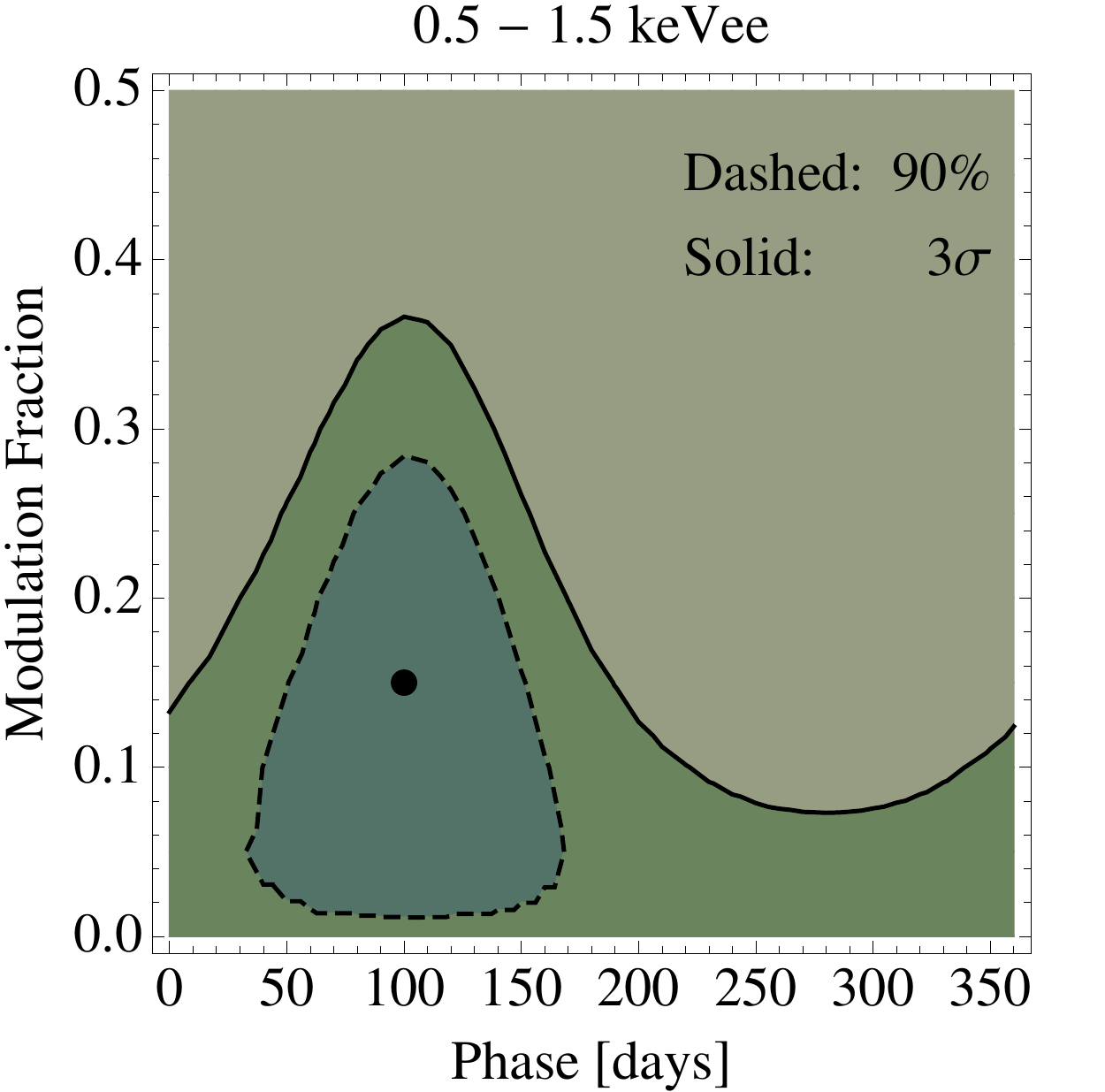}
   \includegraphics[width=2.0in]{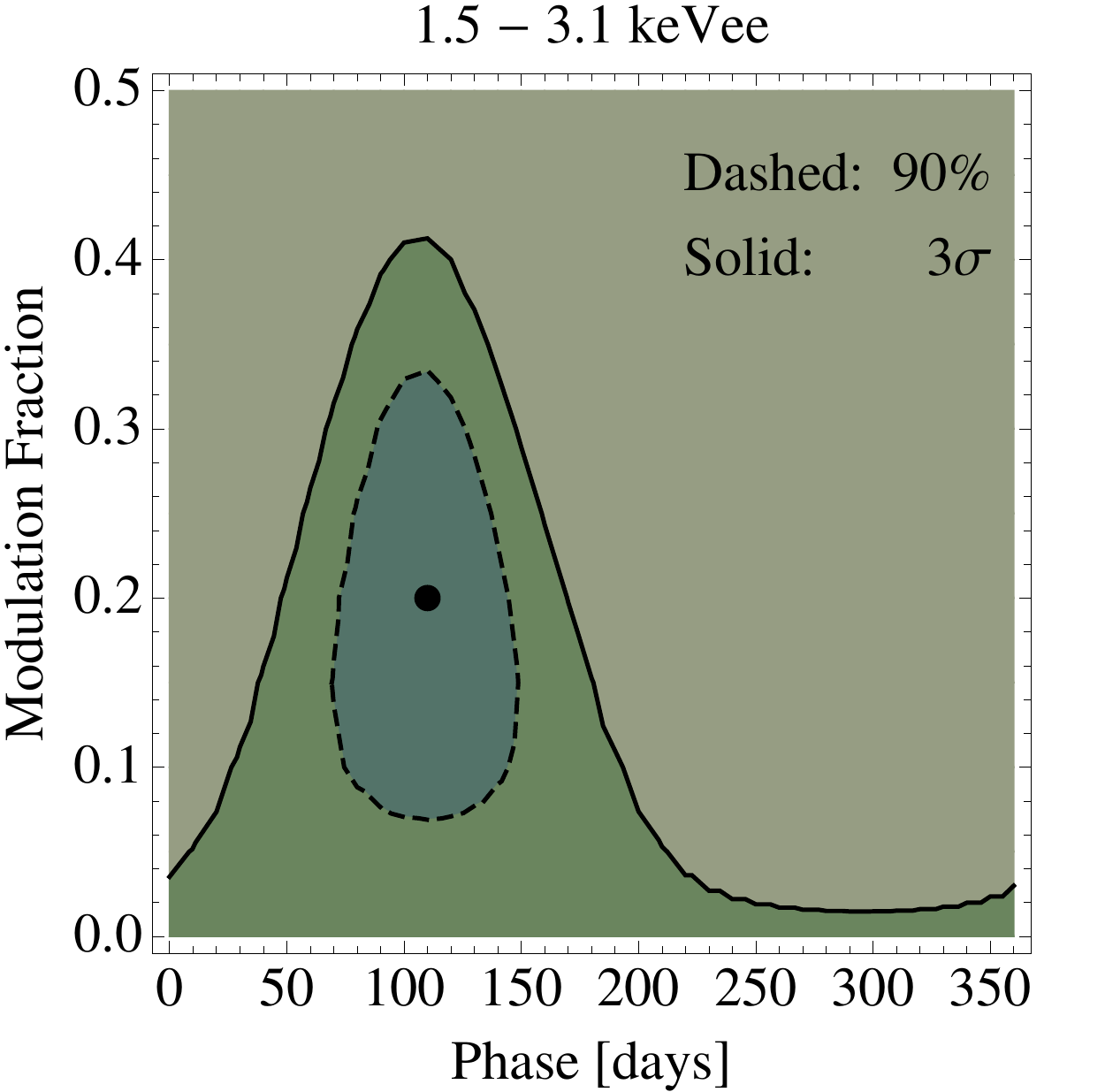} 
 \includegraphics[width=2.0in]{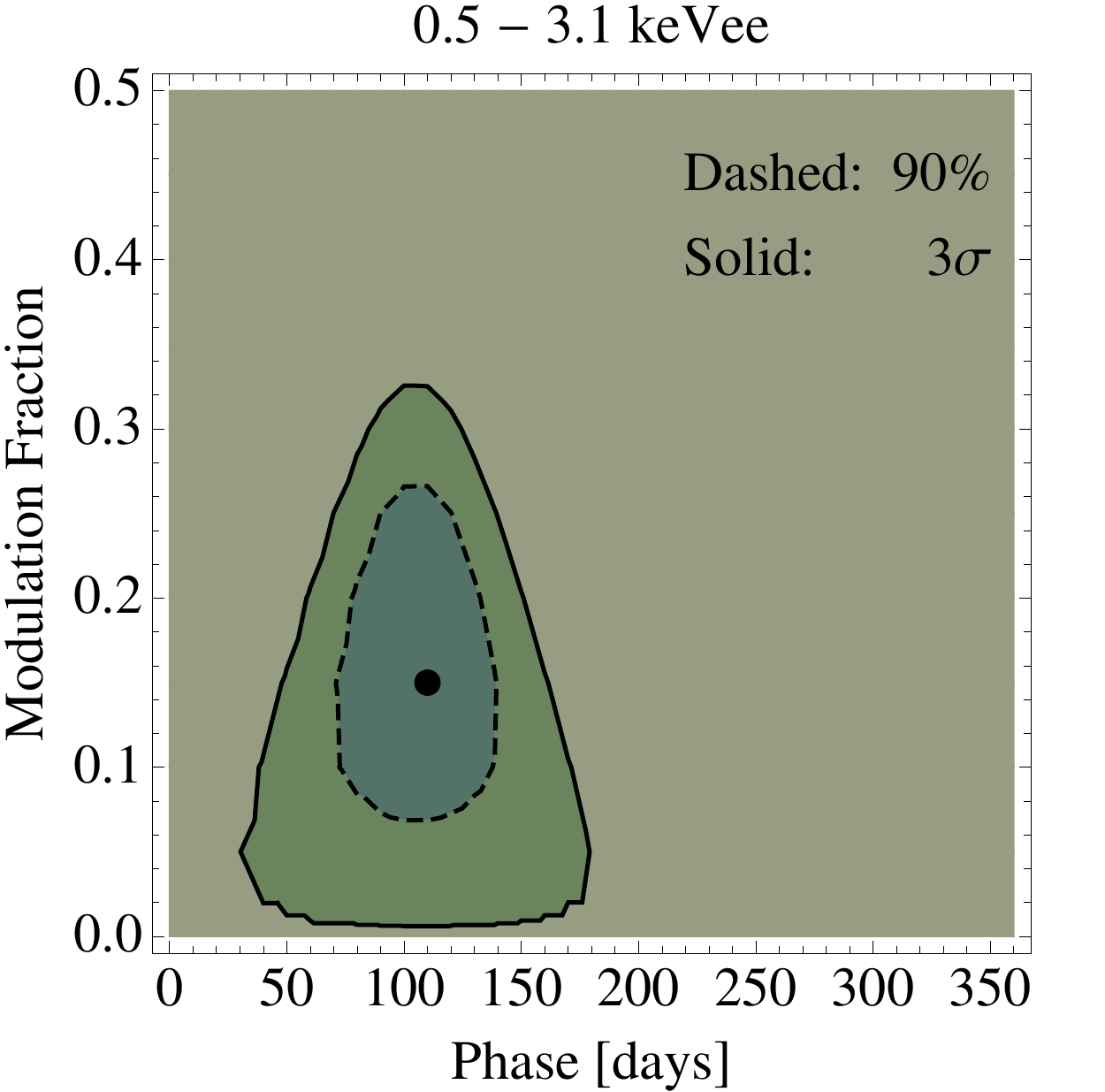}
 \caption{Likelihood analysis of the allowed regions in modulation and phase for different energy ranges: in [0.5--1.5]~keVee ({\it left}), [1.5--3.1]~keVee ({\it middle}), and [0.5--3.1]~keVee ({\it right}).  The contours are of $\Delta\chi^2$  from the best-fit point, shown as $\bullet$.}
   \label{fig:likelihoodregions}
\end{figure}

Next, we consider the phase and amplitude of the modulation for a constant period of one year.  An unbinned (in time) log-likelihood analysis is done for three energy ranges: low [0.5--1.5]~keVee, high [1.5--3.1]~keVee, and all [0.5--3.1]~keVee.  Figure~\ref{fig:likelihoodregions} shows that the high energy data carry nearly the full weight of the analysis, and that the preferred phase is not Maxwellian, confirming the results of the binned analysis in Fig.~\ref{fig:binning}.  The modulation fraction in the high energy range is $\sim 20\%$, with a phase around 106 days.  

\begin{figure}[t]
\begin{center}
  \begin{tabular}{cc}
    \includegraphics[width=6cm]{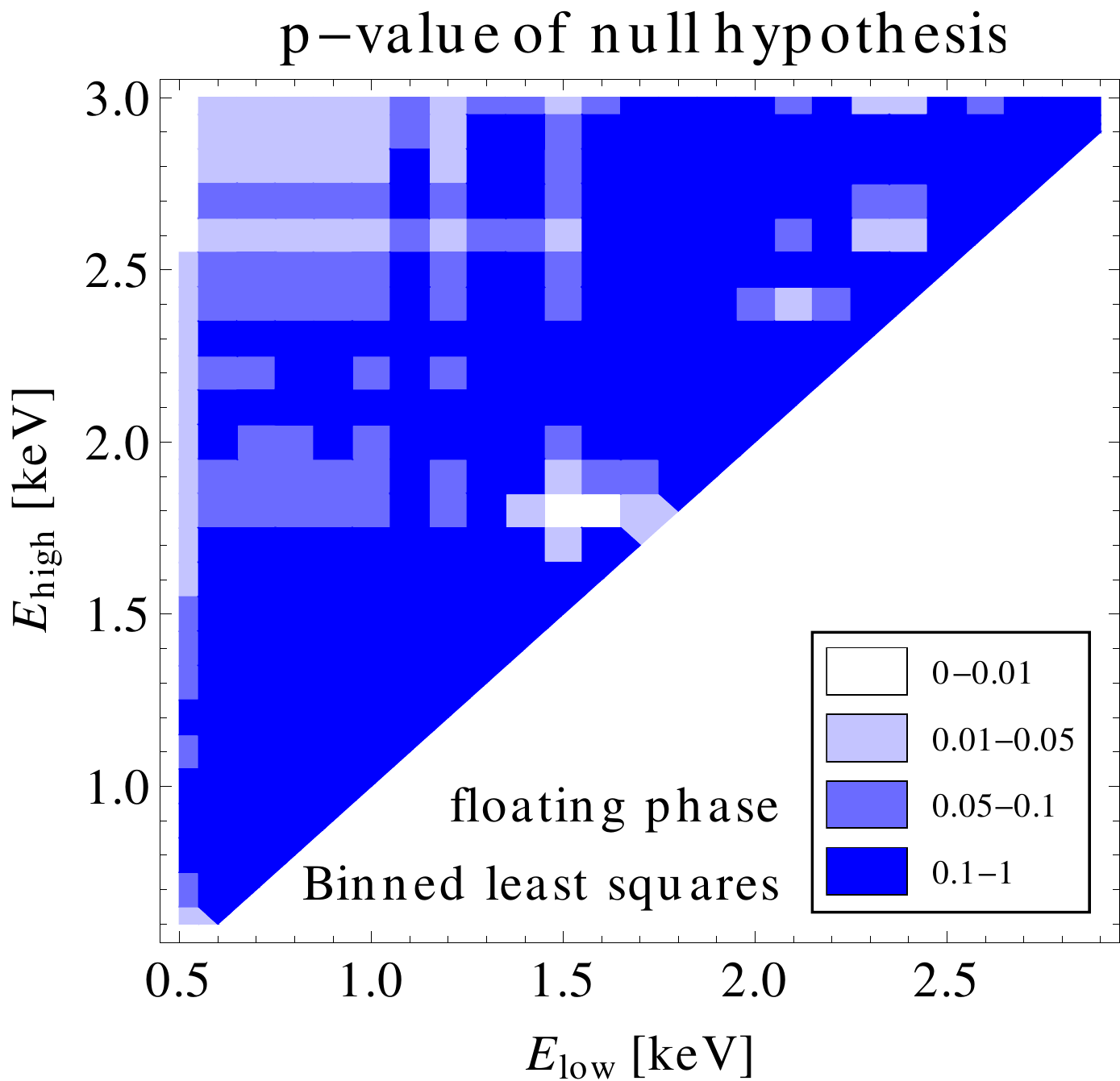} &
    \includegraphics[width=6cm]{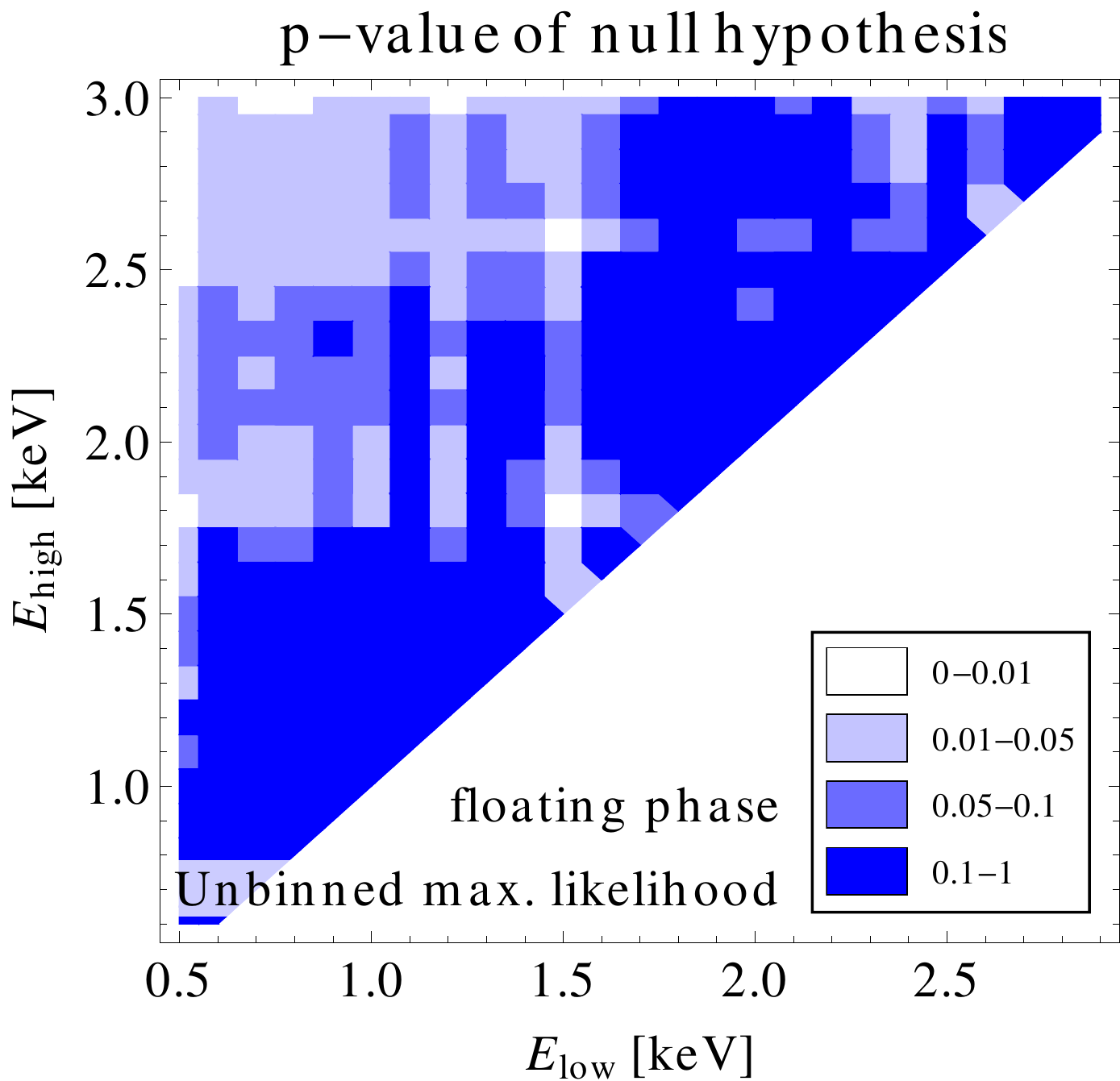}\\
     \includegraphics[width=6cm]{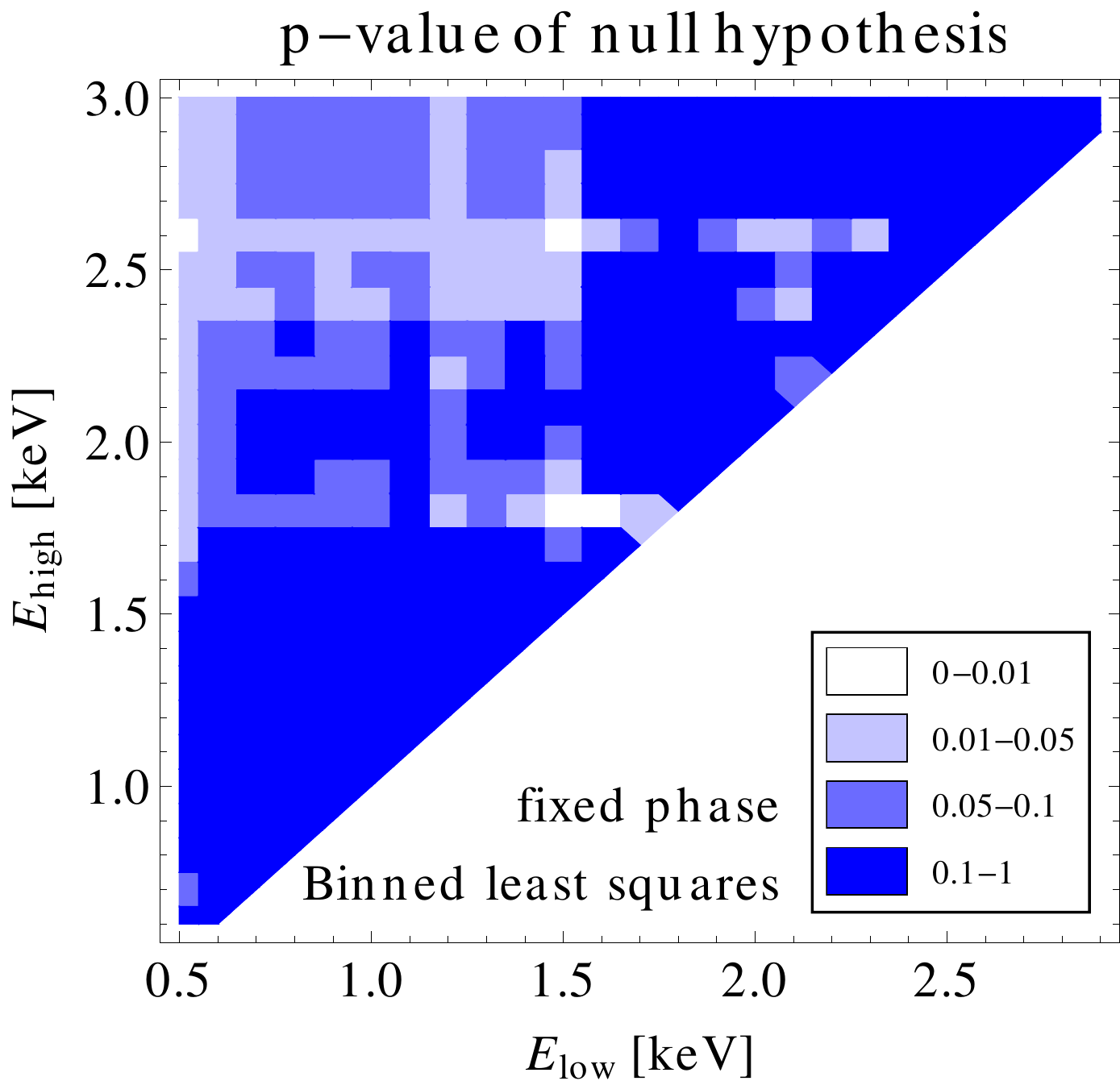} &
    \includegraphics[width=6cm]{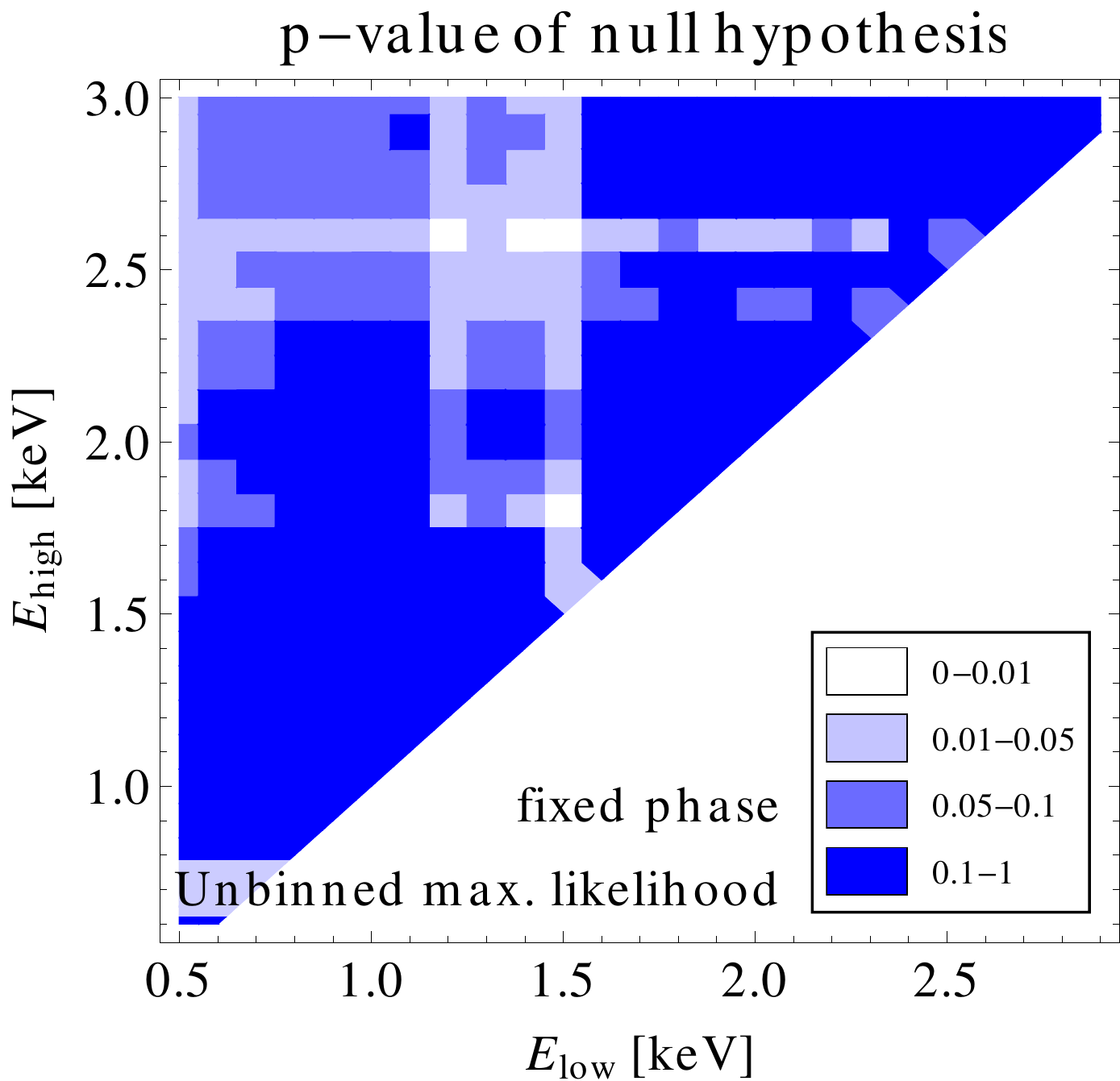}
  \end{tabular}
  \end{center}
      \caption{Results of the modulation fit with $\omega=2\pi /\text{year}$, using both the binned and unbinned approaches. The upper plots allow the phase to float and the lower plots fix it to the value expected for the SHM ($t_0=152$ days). The probability of the null (no modulation) hypothesis to fluctuate to the observed best-fit values is calculated from the $\Delta \chi^2$ between the two best-fits, assuming 2 degrees of freedom for the upper plots and 1 for the lower.}
  \label{fig:At0}
\end{figure}

Figure~\ref{fig:At0} shows the significance of modulation over the null hypothesis in a range of energies from $\elow$ to $\ehigh$, where $\elow, \ehigh$ each go from 0.5 to 3~keVee in steps of 0.1~keVee.  The results for both the binned (left-hand column) and unbinned (right-hand column) analyses are shown to illustrate that the two methods are in very good agreement.  The figure shows the significance when the phase is allowed to float in the upper row, and when the phase is fixed to Maxwellian in the lower row.

The smallest p-values for the null hypothesis occur in the energy range 0.5--3.0~keVee. As in Fig.~\ref{fig:lomb-scargle} (b), there is no significant modulation from $\elow = 0.5$ to $\ehigh \sim 1.7$ keVee, but the significance starts to increase as $\ehigh \gtrsim 1.7$~keVee.  In the energy range where the modulation appears to be most significant, the phase and modulation fraction are both relatively stable; the best-fit phase falls consistently from 60--120 days, while the best-fit modulation fraction falls between 10-20\%, for fits over the full energy range (0.5--3.0 keVee).

 \subsection{Spectra}

\begin{figure}[!t] 
   \centering
   \includegraphics[width=2in]{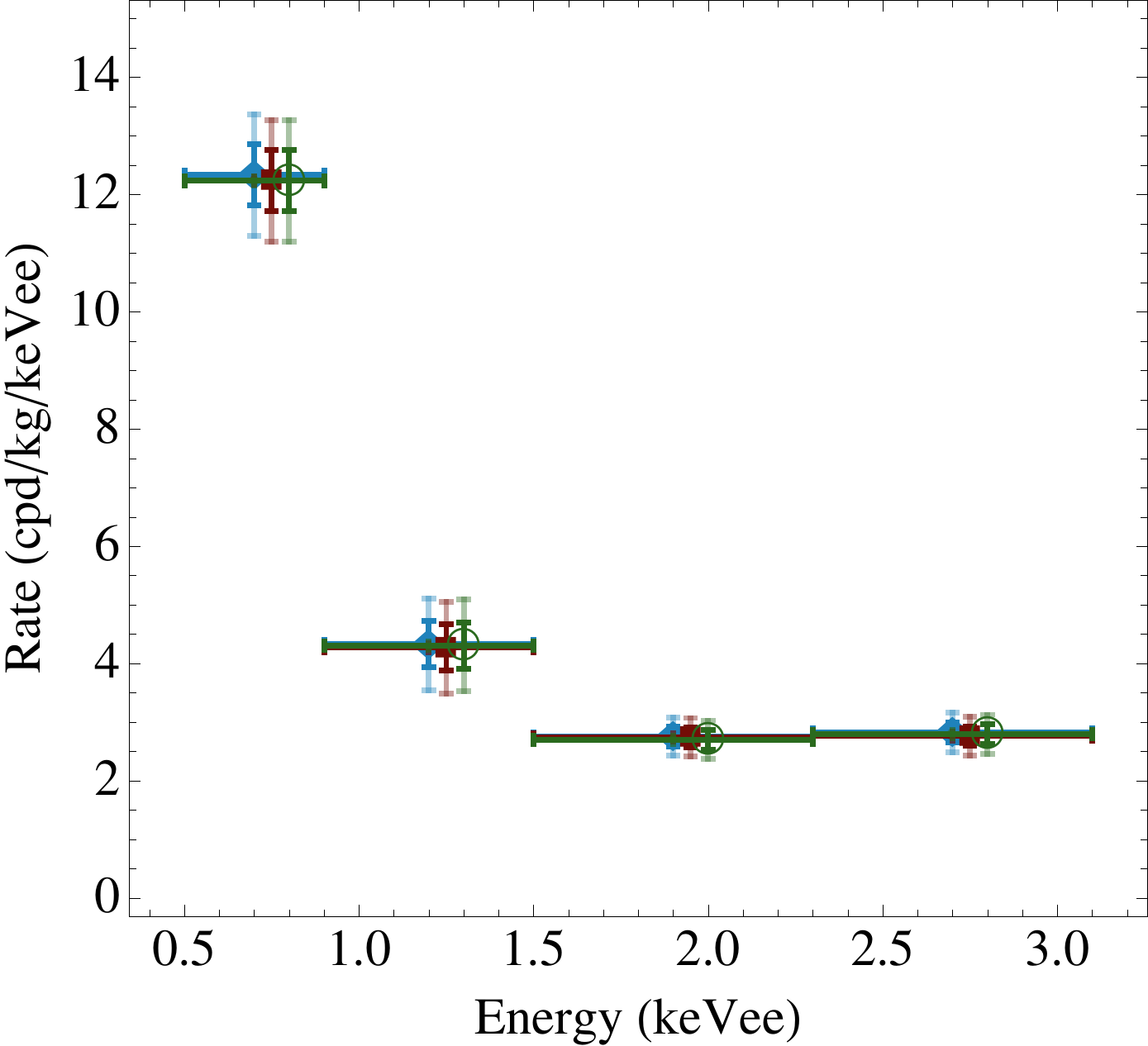} 
   \includegraphics[width=2in]{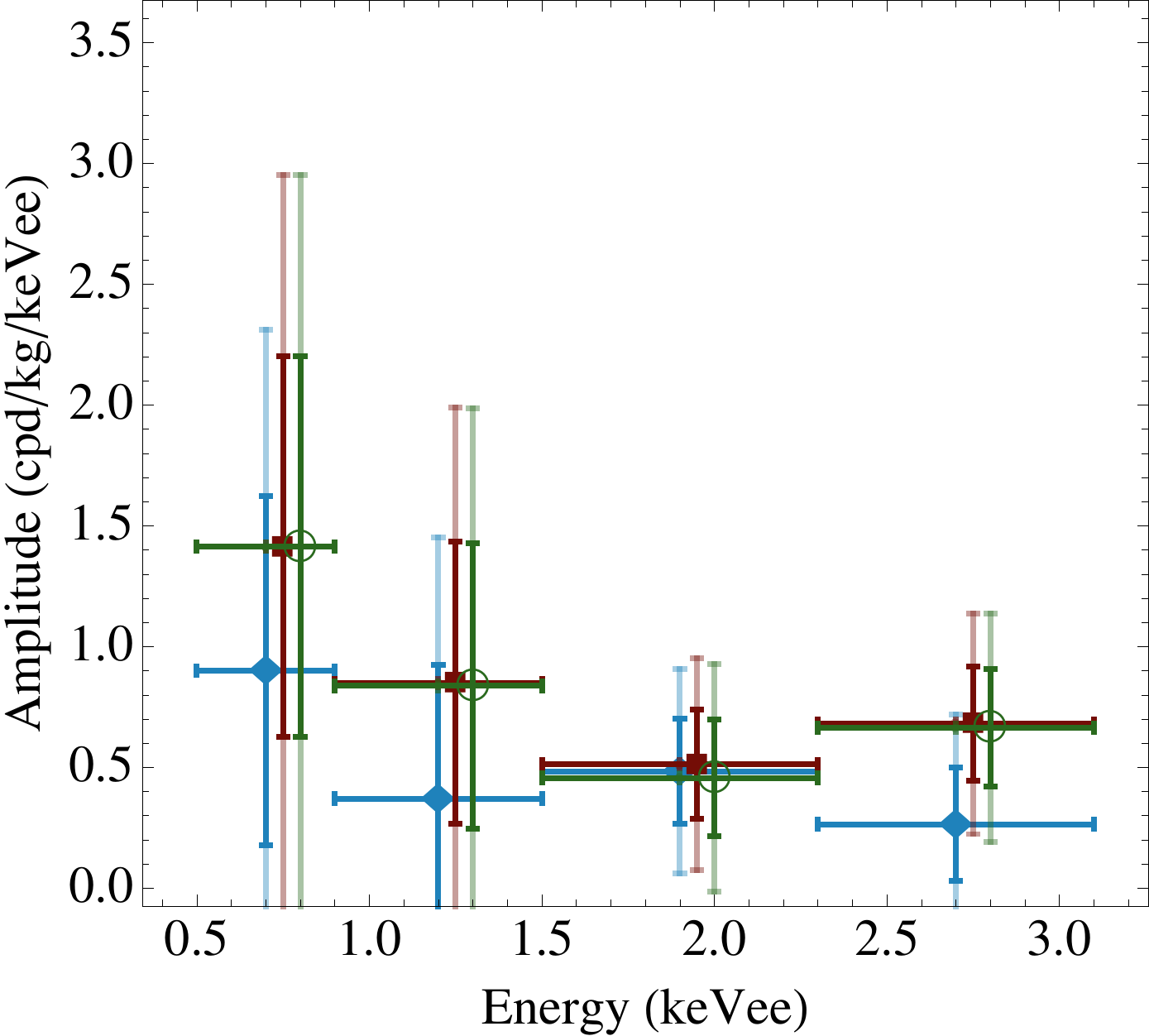}
   \includegraphics[width=2in]{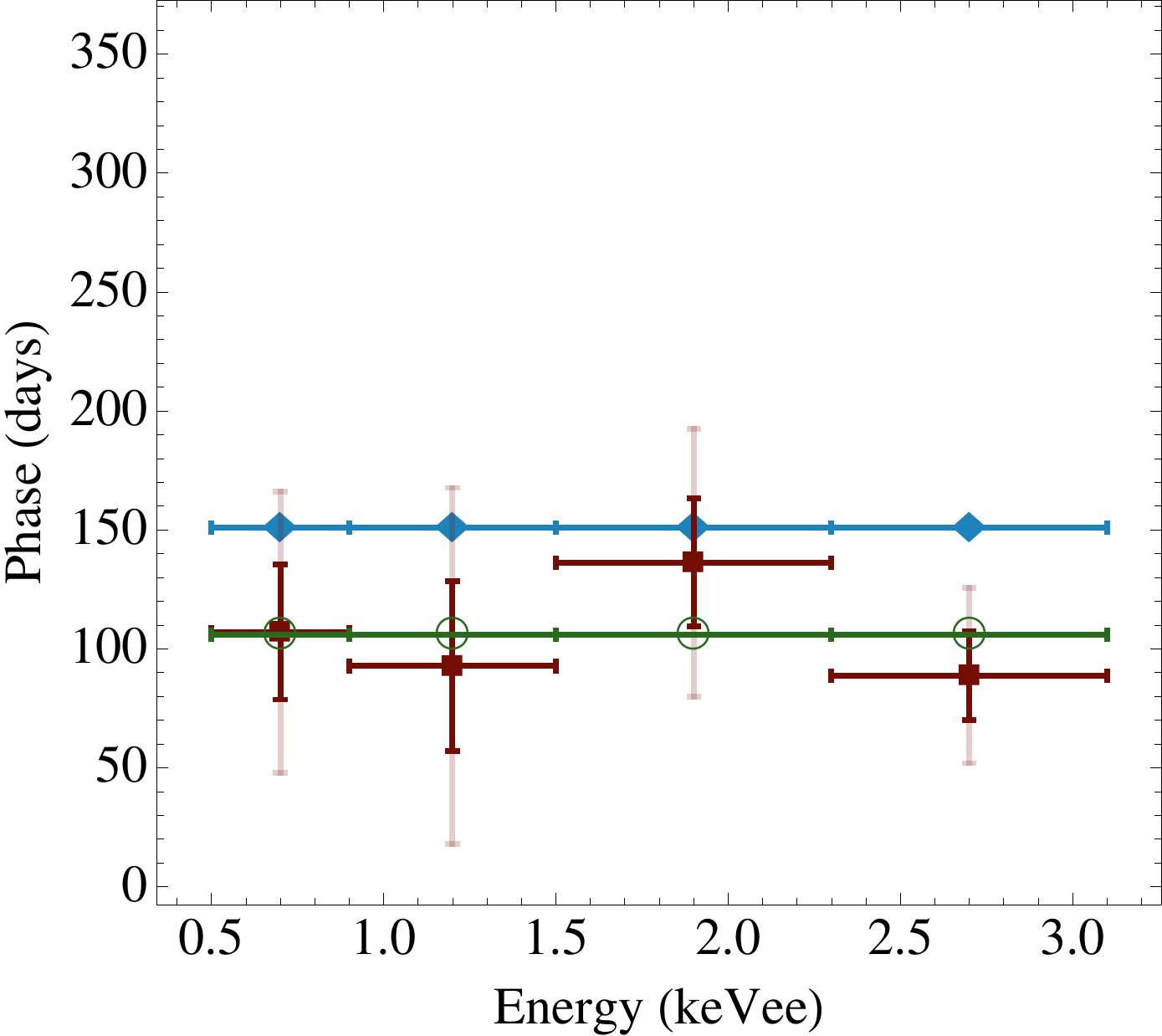}
   \caption{Spectra for three different scenarios:  (red square) the phase is allowed to float in the fit, (blue diamond) the phase is fixed to Maxwellian (152 days), and (green open circle) the phase is fixed to the best-fit phase (106 days) for a fit to the full data range 0.5-3.1 keVee.  The spectra represent:  total rate (left), modulation amplitude (middle) and phase (right).  In all cases the error bars correspond to moving away from the best fit point by $\Delta\chi^2=1$ (dark) or $\Delta\chi^2=4$ (light), the horizontal offsets are for visualization purposes only.}
   \label{fig:LGspectra}
\end{figure}

Having studied the basic properties of the modulation, we now consider the energy spectra of the unmodulated and modulated rate components, and the oscillation phase.  The data are divided into four energy bins: [0.5--0.9], [0.9--1.5], [1.5--2.3] and [2.3--3.1] keVee.  This separates the data into a bin dominantly below the cosmogenic peaks, one encompassing the cosmogenic background, and two evenly spaced bins above the cosmogenic peaks. For each bin, an unbinned (in time) log-likelihood approach is used to obtain the best-fit values for the unmodulated rate ($A_0$), modulated amplitude ($A_0 *A_1$), and phase ($t_0$).

Figure~\ref{fig:LGspectra} shows the best-fit values as a function of energy, once the period is fixed to one year.  Three different scenarios are considered: (blue) the phase is set to Maxwellian ($t_0 = 152$ days), (green) the phase is fixed to the best-fit value for the fit over 0.5-3.1 keVee ($t_0 = 106$ days), and (red) the phase is allowed to float bin-by-bin.  Note that both the unmodulated rates and the modulation amplitudes do not differ dramatically between these three scenarios.  The modulated amplitude has a large value ($\sim 1.5$~cpd/kg/keVee) in the lowest energy bin, but very large error bars ($\pm 0.8$~cpd/kg/keVee).  The modulation amplitude flattens out considerably at higher energies around $\sim 0.5$~cpd/kg/keVee, with smaller error bars.  When the phase is allowed to float, it remains relatively stable over the full energy range, with small variations around the best-fit value.  The significance is highest in the last energy bin for the floating and best-fit phase scenarios; it is largest in the second to last energy bin for the Maxwellian case.
\begin{figure}[!t] 
   \centering
    \includegraphics[width=2in]{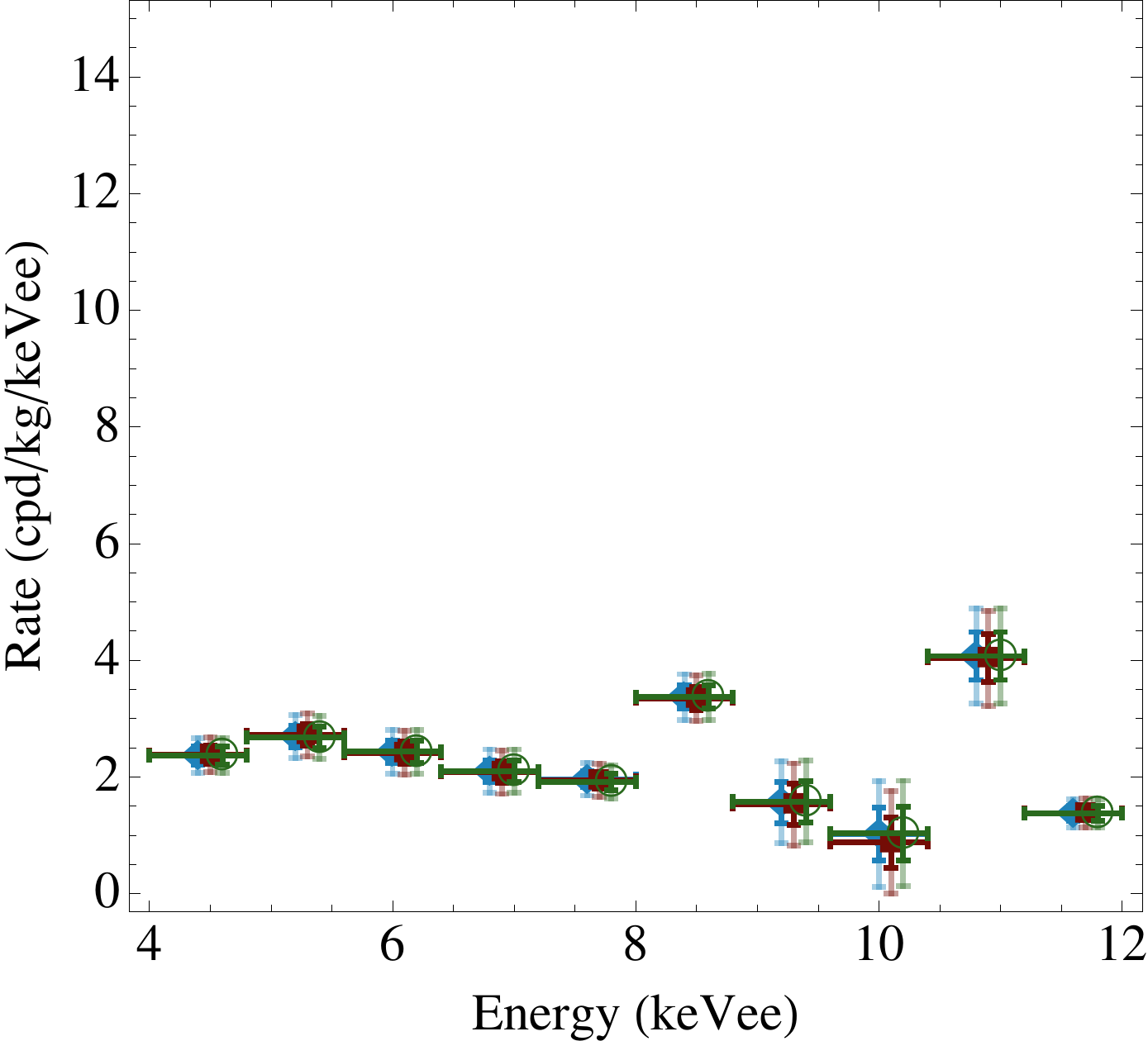} 
   \includegraphics[width=2in]{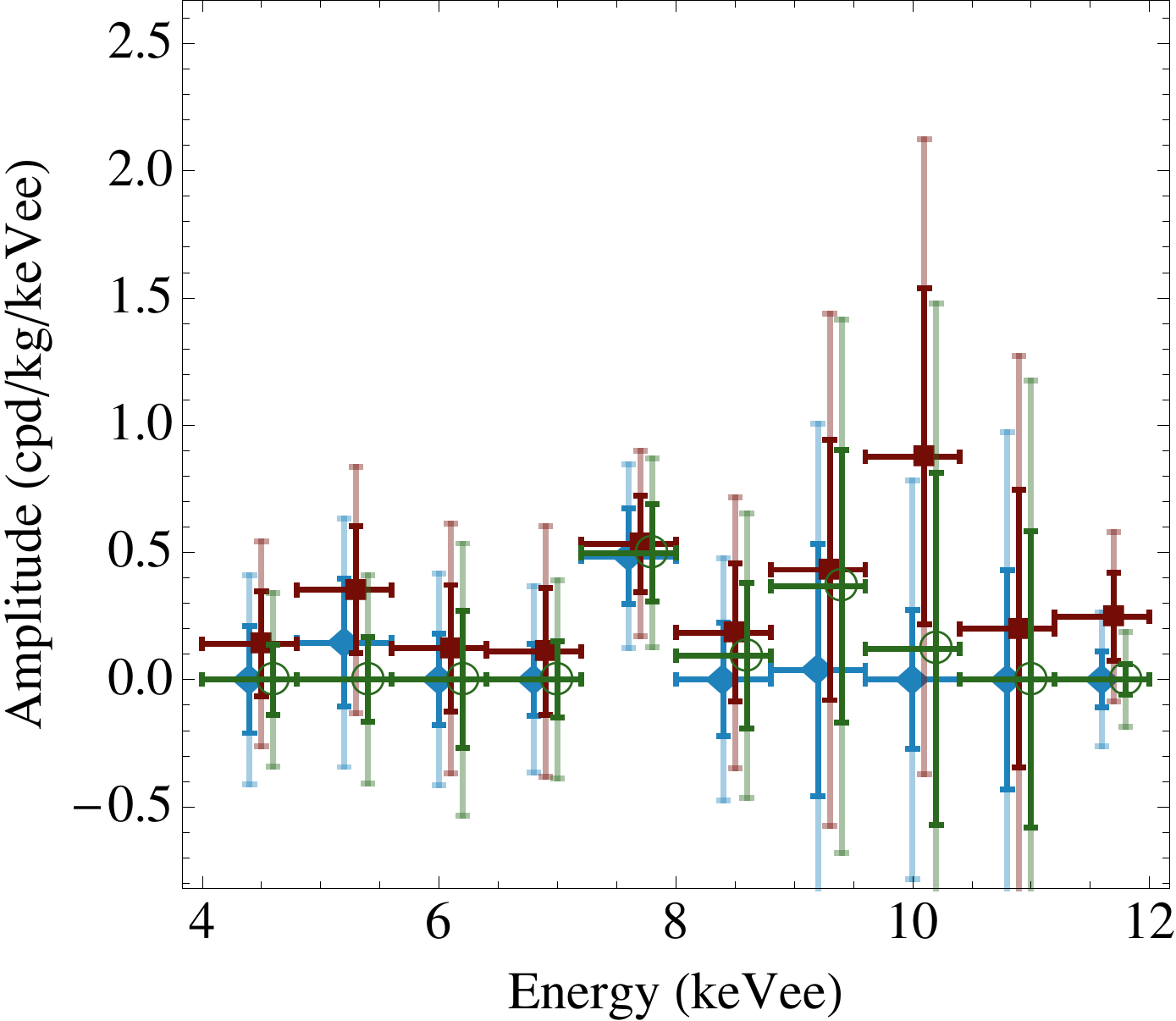}
   \includegraphics[width=2in]{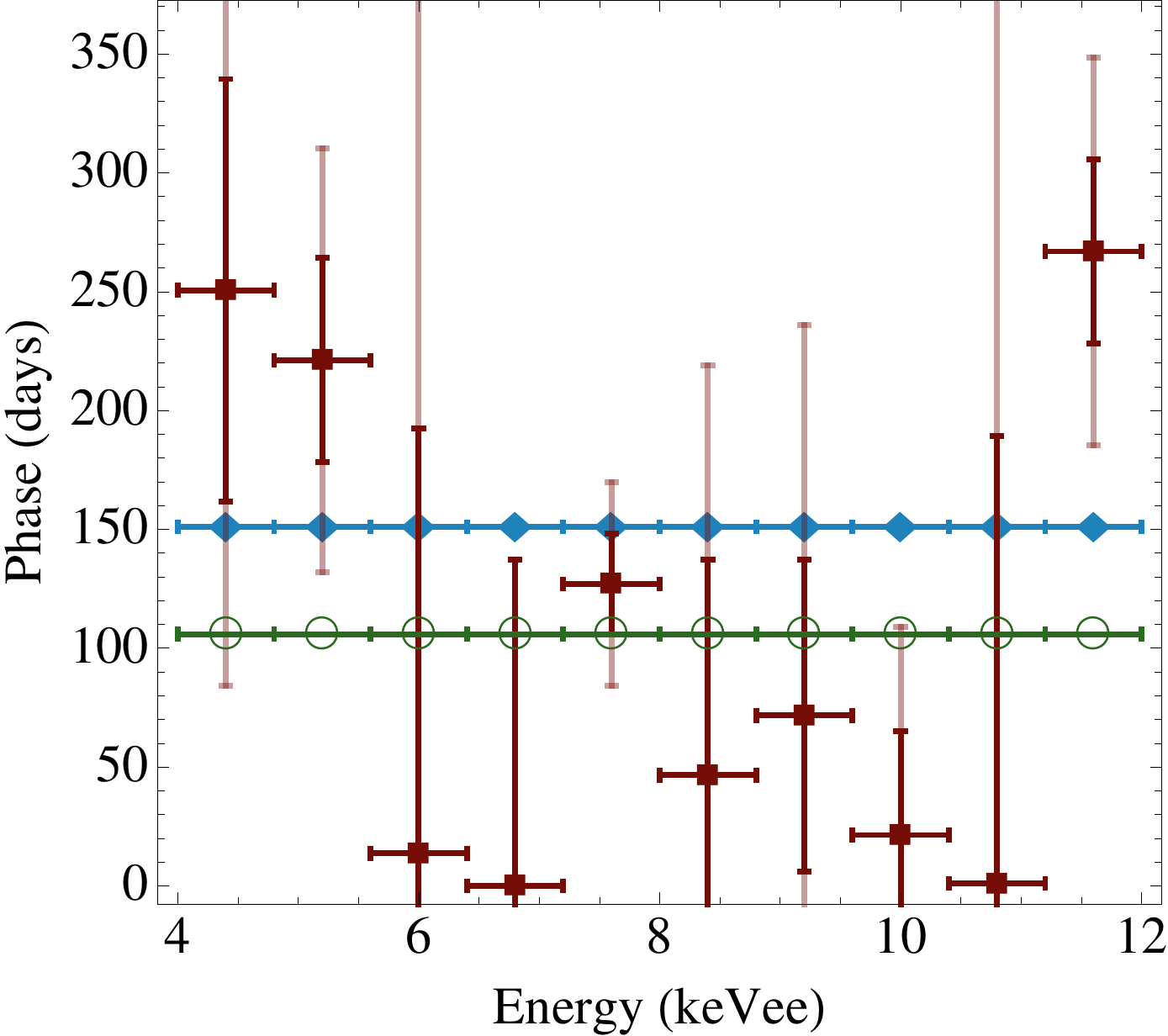}
   \caption{Same as Fig.~\ref{fig:LGspectra}, except for data from the high energy channel.}
   \label{fig:HGspectra}
\end{figure}

The fact that the modulation amplitude in Fig.~\ref{fig:LGspectra} is non-zero even at large energies is a non-trivial and important feature of the spectrum.  To see if this modulation amplitude eventually ``turns off,'' we have also analyzed CoGeNT's high energy channel.  The low and high energy channels in the data both measure pulse amplitudes in the range from 0.05~V and 0.25~V.  However, the relation between the actual physical energy deposited and the measured voltage is different for the two channels.  In particular, the response function at high energies is optimized to provide a good fit to the K-shell cosmogenics from $\sim 4$-12 keVee (see Appendix~\ref{sec:gapsandbacks}), but is not optimized for energies below $\sim$3.2~keVee.  Figure~\ref{fig:HGspectra} shows the results of a log-likelihood analysis in equally-spaced bins above 4 keVee.  The unmodulated rate spectrum indicates that there is an unexplained excess of events that is fairly constant with energy.  In contrast to the low energy channel, the phase is highly unconstrained above $\sim$4 keVee when it is allowed to float in the fitting procedure.  This is due to the fact that there is no signifiant modulation in this energy regime, as illustrated by the spectrum of the modulation amplitude. 

\begin{figure}[t] 
   \centering
    \includegraphics[width=5.5 in]{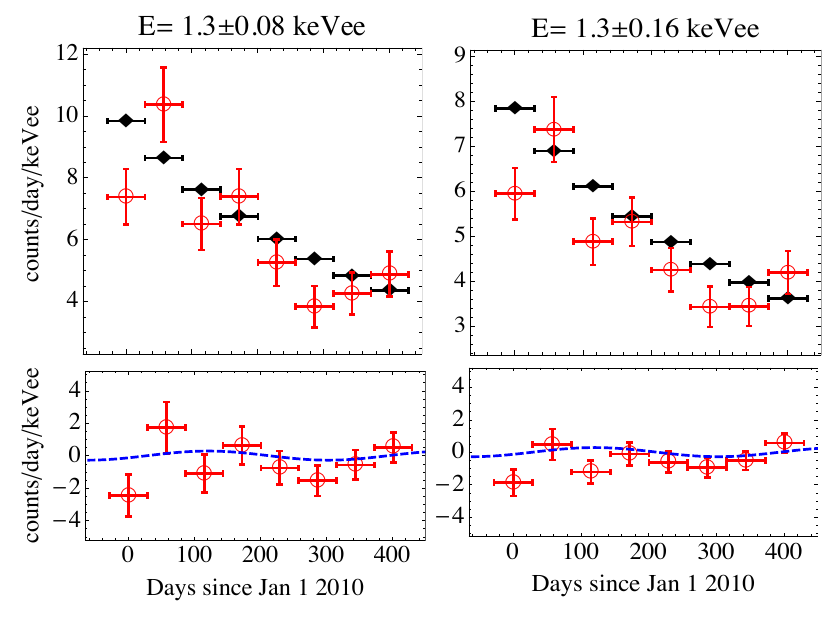} 
    \caption{Time variation for data centered on the ${}^{68}$Ge L-shell peak for two different energy ranges.   The top panel shows the predicted cosmogenic contribution using Eq.~\ref{eq:cosmopeaks} and the parameters given in Appendix~\ref{sec:gapsandbacks} (black diamonds), as well as the (efficiency corrected) time-binned distribution of the data (red open circles).  A constant of 1.4~counts/day/keVee (see unmodulated spectrum in Fig.~\ref{fig:LGspectra}) has been added to the background.  The bottom panel shows the residuals between the data and the model (red).  The dashed blue line is the best-fit modulation in the range 0.9--1.5~keVee, obtained using the log-likelihood approach as in Fig.~\ref{fig:LGspectra}.}
   \label{fig:Bkgdspectra}
\end{figure}

Figure~\ref{fig:LGspectra} shows that there is some modulation in the energy bin from 0.9-1.5 keVee, where the L-shell cosmogenic peaks are expected to dominate.  As a check that the most dominant of these peaks, ${}^{68}$Ge, is not oscillating, we plot the time-binned data for energies centered on this peak in Fig.~\ref{fig:Bkgdspectra} (red open circles).  The black diamonds in the figure show the expected exponentially falling background, modeled using the procedure described in Appendix~\ref{sec:gapsandbacks}.  The bottom panel in the figure shows the residual between the data and model, overlayed with the best-fit modulation in the 0.9-1.5 keVee energy bin from Fig.~\ref{fig:LGspectra}.  As the energy range about $E = 1.3$ keVee is widened, the residuals come into better agreement with the blue line.  In the narrow energy band, the statistics are too small to conclude whether there is a larger modulating component on the peak.  As more data are collected and the statistics improve, it will be crucial to study the time variation of the ${}^{68}$Ge line, especially as with more time it should decay away allowing easier access to any underlying modulation.

\section{A Dark Matter Interpretation}
\label{sec:darkmatter}

\subsection{Dark Matter Fit for Standard Halo Parameters}

The results of the previous section suggest that the distribution of the modulated amplitude is not indicative of a conventional elastically-scattering WIMP.  Furthermore, there is some discrepancy between the best-fit phase and that expected from the standard halo model (SHM).  In this section, we will explore various dark matter (DM) models to see what best fits the observed data.  Unless specified, the velocity distribution of the dark matter is Maxwell-Boltzmann with velocity dispersion $v_0=220\,\kmpers$ and escape velocity $v_{\text{esc}}=550\,\kmpers$:
\be
f(v)\propto (e^{-v^2/v_0^2}-e^{-v_{esc}^2/v_0^2}) \, \Theta(v_{esc}-v)~,
\ee
where $v$ is the velocity in the galactic rest frame.  More general velocity profiles will be considered in the following subsections.

We carry out fits using an unbinned extended maximum likelihood approach and a binned $\chi^2$ analysis. For the unbinned method, we define a likelihood function that includes the dark matter signal, the cosmogenic backgrounds, and a constant background with floating normalization. The likelihood function accounts for efficiencies and shutdown periods. For the binned approach, the data is divided into five energy bins of equal size, spanning the range from 0.5--3.0~keVee. Within each energy bin, the events are partitioned in fifteen equal-sized time bins, each approximately one month wide.  We subtract cosmogenic backgrounds and correct for the shutdown periods of the detector (the efficiencies are accounted for in the predicted dark matter signal). The error in each bin is based on the statistical uncertainty before background subtraction and deadtime correction, and is assumed to be approximately Gaussian, which is a reasonable assumption because no bin has fewer than six events.  We carry out a $\chi^2$ fit to these 75 bins, minimizing over dark matter parameters and systematic nuisance parameters. 

\begin{figure}
  \begin{center}
    \includegraphics[width=0.6\textwidth]{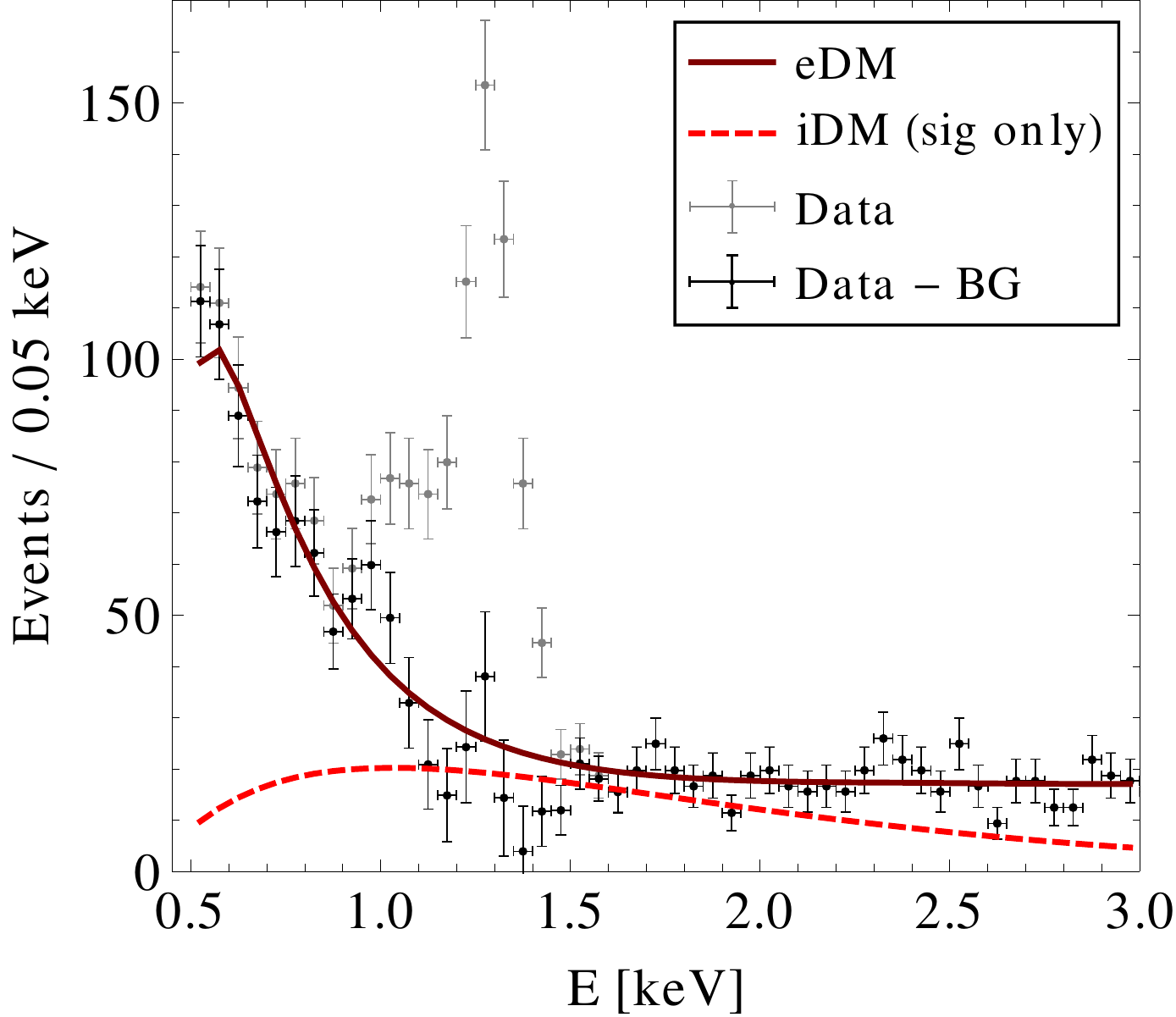}
  \end{center}
  \caption{Comparison of the CoGeNT data to the predicted event spectrum for elastically (solid dark red) and inelastically (light red dashed) scattering dark matter. In both cases, results for the best-fit dark matter parameter are shown. In the elastic case, the fit is done using the unbinned maximum likelihood approach, including energy and timing information for each event as well as a constant background.  For the inelastic case, a binned analysis is done with a background that can float in each of five energy bins.  (For the iDM case, the signal is shown, but not the fitted constant background.)}
  \label{fig:bf-spectrum}
\end{figure}

\begin{figure}
\begin{center}
\includegraphics[width=\textwidth]{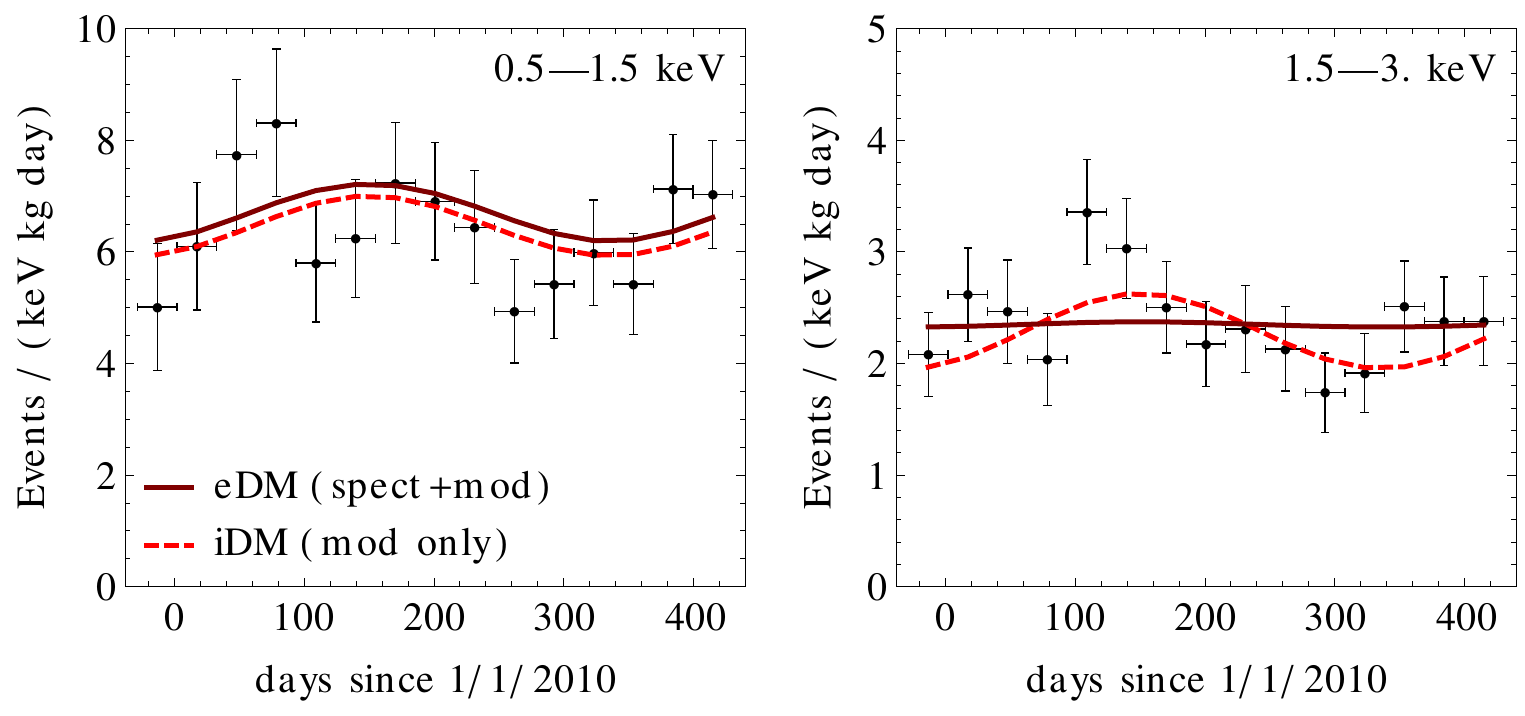}\\
\caption{Comparison of the CoGeNT data to the predicted time-dependent event rate for elastically (solid dark red) and inelastically (light red dashed) scattering dark matter, using the same fitting procedures as for Fig.~\ref{fig:bf-spectrum}.}
\label{fig:bf}
\end{center}
\end{figure}

Several different signal and background scenarios are considered and the significance for each is summarized in Table~\ref{tab:eDMtab}.  These results should be compared to a background-only fit where a constant rate is assumed in each of the five energy bins with no time variation.  This fit gives $\chi^2= 58.2 $ for 70 degrees of freedom (d.o.f.).  The separate scenarios are:
\begin{itemize}

\item \textbf{Spectrum + Modulation}
We attempt fitting elastic DM (eDM), varying $\sigma$ and $m_\chi$, to both the energy spectrum and its time dependence.  We assume an additional background contribution, constant in time and energy, and include its rate as a nuisance parameter, $c_0$.  The binned analysis gives $\chi^2 = 57.3$ for 72 d.o.f..  The predicted and observed rates for the best-fit point obtained from the unbinned method is also shown in Fig.~\ref{fig:bf-spectrum} and \ref{fig:bf}.  The best-fit for the eDM scenario, which has a mass $\sim 7$ GeV, is marginally better than the time-independent background-only fit; the corresponding p-value is 0.64.  

\item \textbf{Spectrum Only}
The second column of Table~\ref{tab:eDMtab} shows an eDM scenario with a constant background, but this time fitting only to the unmodulated spectrum, i.e.\ ignoring time information.  In this case, the binned least-squares analysis uses 50 bins and gives $\chi^2=50.8$ for 47 d.o.f.

\item \textbf{Modulation Only}
The third and fourth columns of Table~\ref{tab:eDMtab} only fit to the modulation in the data.  The fits are done for an eDM plus background hypothesis, where the time-independent background is allowed to vary freely in each of the five energy bins (i.e., there are five nuisance parameters $c_i$).  For the case where these constant contributions, presumably coming from some unidentified background, are restricted to be physical ($c_i \ge 0$), the fit gives $\chi^2 = 53.7$ for 68 d.o.f.  If instead they are allowed to float, $\chi^2 = 51.4$, again with 68 d.o.f.  Notice that when fitting the modulation, the best-fit dark matter mass is $\sim 10-12$ GeV.

\end{itemize}

\begin{table}
\begin{center}
\begin{tabular}{cccccccc}
\hline
Scenario                 &\multicolumn{2}{c}{Spect+Mod}
                                       &\multicolumn{2}{c}{Spect only}
                                                      & Mod  only    & Mod only  & iDM  \\
                         &    &        &      &       & ($c_i\ge 0$) &           & Mod only  \\
\hline
d.o.f.                   & 72  & [n/a] & 47   & [n/a] & 68           & 68        & 67   \\
$\sigma/10^{-41}$ cm$^2$ & 13.8& [8.9] & 10.1 & [8.2] & 6.0          & 8.6       & 64   \\
$m_\chi/\gev$            & 7.2 & [8.1] & 7.7  & [8.2] & 10.0         & 12.0      & 16.3 \\
$\delta/\kev$            &     &       &      &       &              &           & 24   \\
$c_i/$(cpd/kg/keVee)    & 2.5 & [2.5] & 2.5  & [2.6] & $\begin{pmatrix} 1.5 \\ 0 \\ 1.5 \\ 2.3 \\ 2.3 \end{pmatrix}$  & $\begin{pmatrix} -5.9 \\ -4.6 \\ -1.0 \\ 1.1 \\ 1.7 \end{pmatrix}$ & $\begin{pmatrix} 8.7 \\ 0.4 \\ 0.4 \\ 1.2 \\ 1.5 \end{pmatrix}$\\
\hline
$\chi^2$                 & 57.3 &[n/a] & 50.8 & [n/a] & 53.7         & 51.4      & 51.3 \\
\hline
\end{tabular}
\end{center}
\caption{The best-fit dark matter parameters for the binned [unbinned] analyses of various DM scenarios.  The numbers should be compared to a background-only fit that gives $\chi^2= 58.2 $ for the 70 degrees of freedom (see text for details).}
\label{tab:eDMtab}
\end{table}

\begin{figure}
  \begin{center}
  \includegraphics[width=0.6 \textwidth]{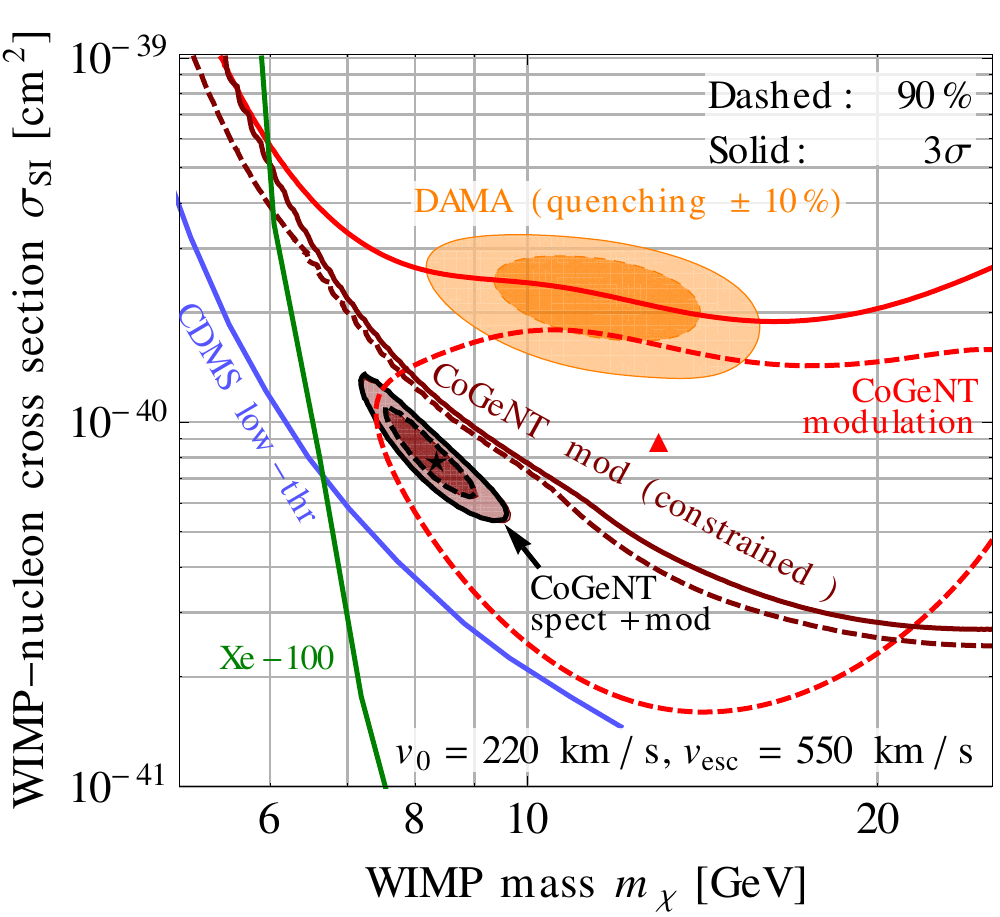} 
  \caption{Preferred regions and exclusion limits at 90\% and $3\sigma$ confidence
    level in the $m_\chi$--$\sigma$ plane for spin-independent dark
    matter--nucleon scattering assuming a standard Maxwell-Boltzmann halo with
    escape velocity $v_{\rm esc} = 550$~km/s and velocity dispersion $v_0 =
    220$~km/s. Filled red (dark gray in B/W) contours are obtained from an unbinned maximum
    likelihood fit to the CoGeNT data, using both the energy and timing
    information for each event. (A fit using only energy information gives
    practically identical results.) The unfilled red (gray) contours are from a binned
    $\chi^2$ analysis, using only the timing information and leaving the energy
    spectrum completely unconstrained (light red/light gray contours), or requiring the
    predicted energy spectrum to remain below the observed one (dark red/dark gray
    exlcusion limits). The orange (light gray) region shows the masses and cross sections
    preferred by DAMA~\cite{Bernabei:2008yi} if the quenching factors are assigned a 10\%
    uncertainty~\cite{Kopp:2009qt,Hooper:2010uy,Belli:2011kw}, and the blue and
    green contours indicate the 90\% exclusion limits from
    CDMS~\cite{Ahmed:2010wy} and XENON100~\cite{Aprile:2011hi}, respectively.}
  \label{fig:eDMfig}
  \end{center}
\end{figure}

Table~\ref{tab:eDMtab} shows that although the data are consistent with the SHM DM hypothesis, the inclusion of such a DM component does not greatly improve the fit compared to the hypothesis of a background that is constant in time.  Furthermore, there is slight tension in the DM interpretation---the modulation in the data by itself favors heavier dark matter masses than the spectrum. This behavior is also illustrated in Figure~\ref{fig:eDMfig}, where we show the preferred region in DM mass $m_\chi$ and elastic spin-independent scattering cross section $\sigma_{\rm SI}$ for various fits to the CoGeNT data.  The plot confirms the slight (but not statistically significant) tension between the DM masses preferred by the energy spectrum observed in CoGeNT and the annual modulation.  A fit to the modulation data alone can exclude the hypothesis of no DM at low confidence level (light red contours), however this requires an unphysical background  (i.e., $c_i \leq 0$).
If positive background is required, the modulation provides only an upper bound. We also compare the CoGeNT-preferred regions to the exclusion limits from XENON100~\cite{Aprile:2011hi} and CDMS~\cite{Ahmed:2010wy}, and confirm the well-known tension between these results.  To compute the DM parameter region favored by DAMA~\cite{Bernabei:2008yi}, we follow the procedure described in~\cite{Kopp:2009qt}, assuming no channeling~\cite{Bernabei:2007hw,Bozorgnia:2010xy} and assigning a 10\% systematic uncertainty to the DAMA quenching factors~\cite{Bernabei:1996vj,Hooper:2010uy,Belli:2011kw}. This systematic uncertainty has a strong effect on the horizontal extent of the DAMA region.  For standard halo parameters, the DM interpretations of DAMA and CoGeNT are inconsistent.  To some extent, this tension can be relaxed for halo parameters different from the ones we chose in Fig.~\ref{fig:eDMfig}~\cite{McCabe:2010zh,Lisanti:2010qx}. In addition, one might speculate that systematic uncertainties in either DAMA or CoGeNT can be larger than what was assumed here.  For instance, if the low-energy event excess in CoGeNT is only partly due to dark matter and partly due to some other source, higher DM masses may become allowed.

Because an elastically scattering WIMP with a Maxwell-Boltzmann velocity distribution only produces a large modulation amplitude at low energies ($\lesssim 1.5$ keVee), it is worthwhile to consider non-standard WIMP scenarios as an explanation of the CoGeNT modulation.  Inelastic dark matter (iDM) is an example of a DM scenario with increased modulation and a preference for events at high recoil energy~\cite{TuckerSmith:2001hy}.  We can repeat the exercise above under the assumption that DM scatters inelastically~\cite{Frandsen:2011ts}.  However, there is no preference for iDM from fits to the unmodulated spectrum, which has many events below where an inelastic contribution is expected - see  Fig.~\ref{fig:bf-spectrum}.  There is, however, a preference for iDM if one fits to only the modulation (see last column in Table~\ref{tab:eDMtab}).  Interestingly, for the iDM hypothesis, there is no improvement in the fit if the background is allowed to be unphysical ($c_i < 0$).  Instead, the iDM model enables a similar modulation spectrum as the eDM fit, but with physical backgrounds.  For the modulation-only fit, iDM gives $\chi^2 = 51.3$ for 67 d.o.f. The spectra for this best-fit point are shown in Fig.~\ref{fig:bf}; clearly, iDM can produce modulation in all energy bins, but at the cost of explaining only part of the event excess observed by CoGeNT.  We note that inelastic WIMPs are highly sensitive to non-Maxwellian properties. If the phase is truly shifted away from the Maxwellian 152 day peak, this could be indicative of such halo properties.  The CoGeNT best-fit phase is mid-April, during the time when XENON100 had increased levels of noise; therefore, a signal localized in this time could have been missed.  Because iDM in the presence of a stream can lead to narrow peaks for brief periods of the year \cite{Lang:2010cd}, further running of XENON100 in April should clarify this situation.

\subsection{Varying the Halo Parameters}

Next, we explore whether the CoGeNT data are compatible with a general class of equilibrium velocity distributions that extend beyond Maxwell-Boltzmann.  In particular, we consider distributions of the form 
\begin{equation}
f(v) \propto (e^{-v^2/v_0^2} - e^{-v_{\text{esc}}^2/v_0^2})^k \, \Theta(v_{esc}-v) \,,
\label{eq: nfwhalo}
\end{equation}
where $k$ is a power-law index, $v_{\text{esc}}$ is the escape velocity, and $v_0$ is the dispersion.  Note that $k=1$ is just the Maxwell-Boltzmann-like halo.  This velocity distribution models the behavior of double power-law density profiles and corresponds to results found in high-resolution simulations of the Galactic halo, when $k\sim2$~\cite{Lisanti:2010qx}.  The fact that simulations support a power-law index greater than one suggests that the number of high velocity particles on the tail of the distribution may be suppressed relative to the expectation for Maxwell-Boltzmann halos.
\begin{figure}[tb] 
   \centering
   \includegraphics[width=2.75in]{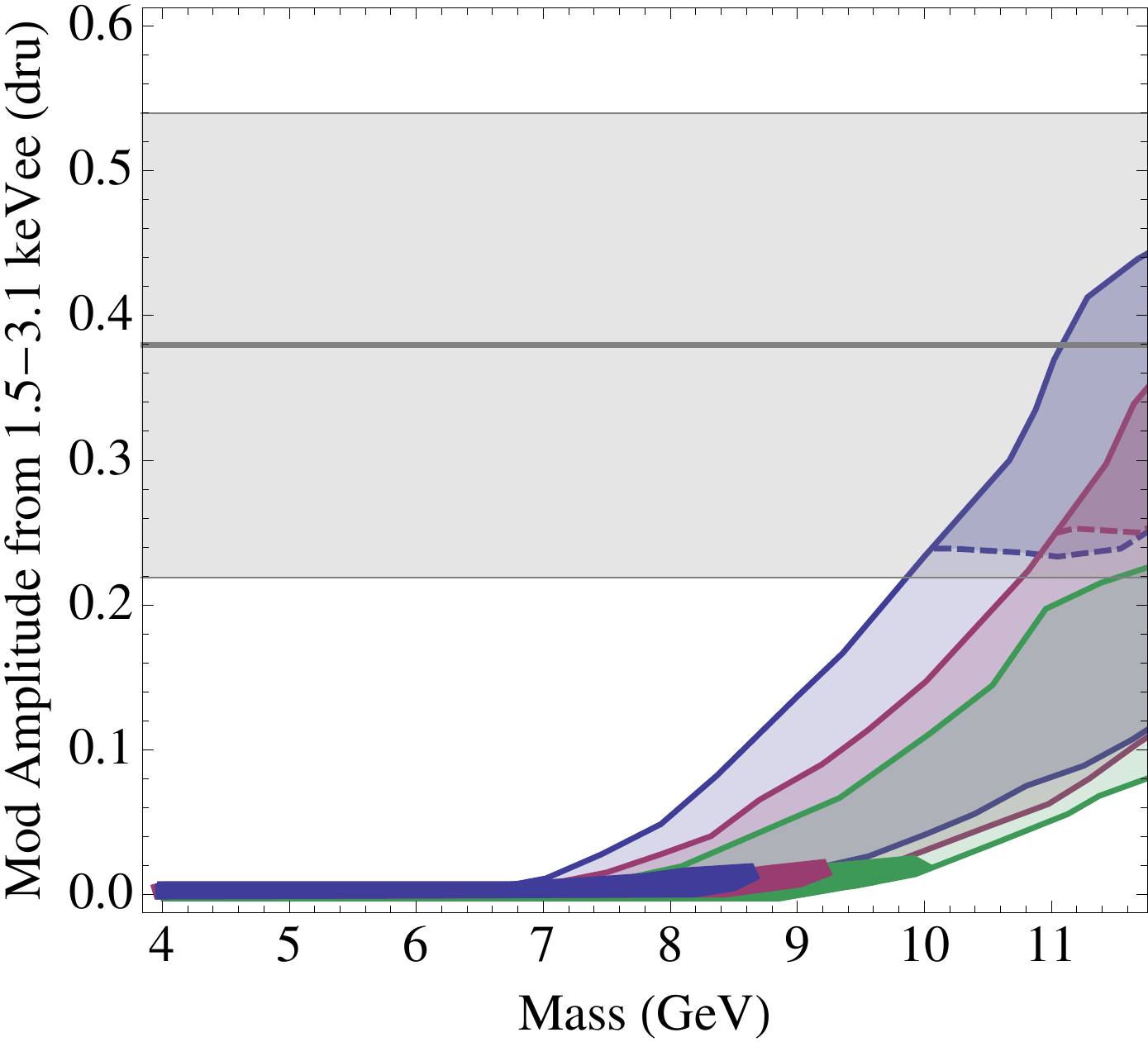}
   \includegraphics[width=2.75in]{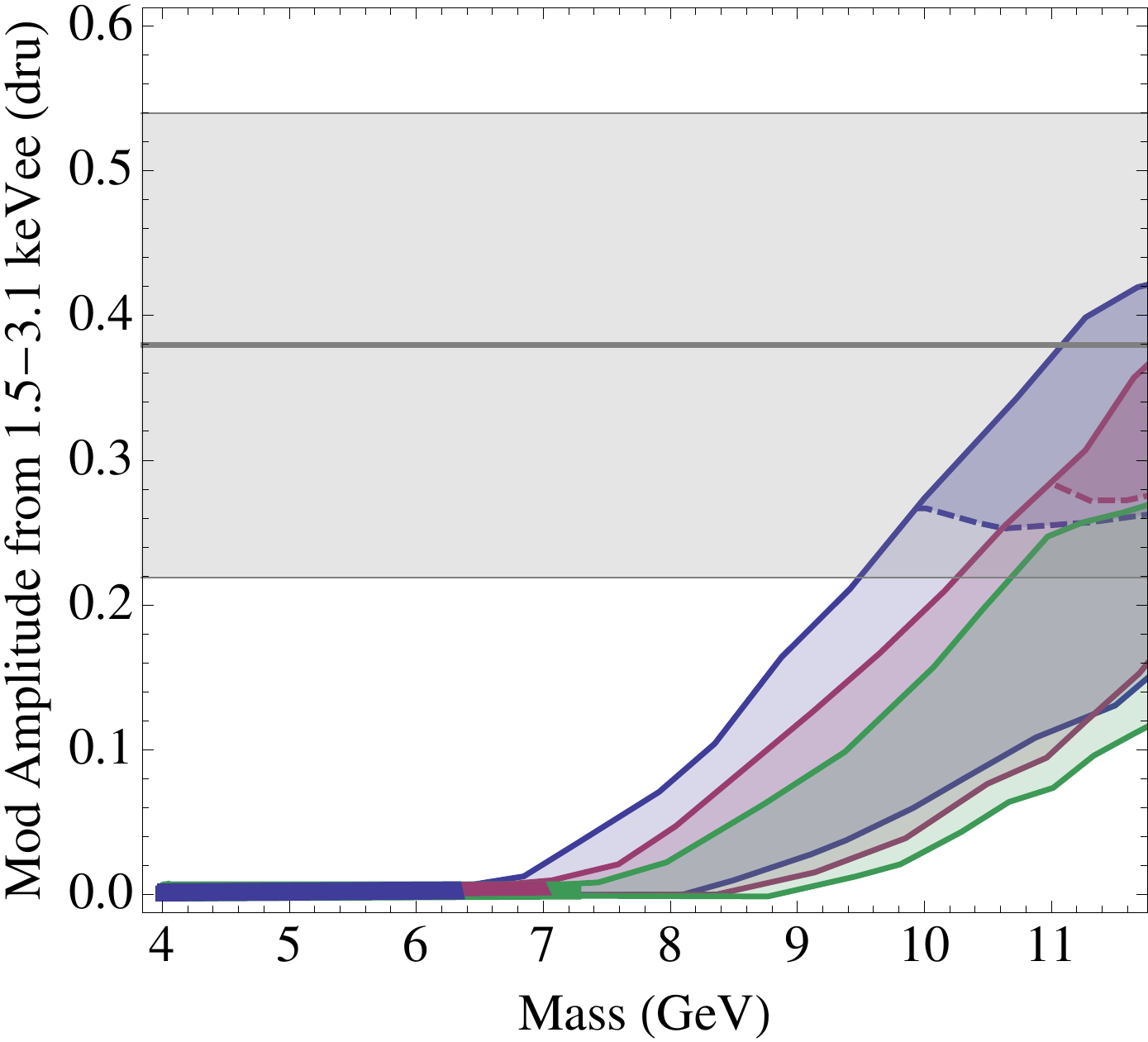}  
   \caption{Modulation amplitude in the range 1.5--3.1~keVee as a function of dark matter mass, where the dark matter cross section is normalized to fit the modulation amplitude in the first bin (left) and over the whole energy range (right).  The colors indicate different spectral indices for Eq.~\ref{eq: nfwhalo}: k=1 (blue), k=2 (pink), k=3 (green).  The regions between (above) the solid (dashed) lines indicate points that overpredict the unmodulated rate by at least 2$\sigma$ from 0.5--1.5 keVee (1.5--3.1 keVee).  The solid colored bands are the only regions consistent with the unmodulated rate spectrum.  The gray band is the modulated amplitude with 1$\sigma$ error bars for the 1.5--3.1~keVee region.}
   \label{fig:nfwplots}
\end{figure}

To study how the CoGeNT predictions are affected by variations in the halo parameters, distributions with $k=1,2,3$ are considered and a random scan is done with $v_{\text{esc}} \in [500,600]$ km/s~\cite{Smith:2006ym} and $v_0 \in [180,280]$ km/s~\cite{Bovy:2009dr, McMillan:2009yr, Reid:2009nj}.  For each randomly selected set of halo parameters and dark matter mass, the modulated and unmodulated rates are evaluated in two energy bins: 0.5-1.5 keVee and 1.5 - 3.1 keVee.  The results are summarized in Fig.~\ref{fig:nfwplots} for the case where the cross section is normalized to give the measured modulation amplitude in the first bin (left) and the case where it is normalized to the total amplitude from 0.5-3.1 keVee (right).  The three colors represent three different spectral indices, with k=1 Maxwellian-type in blue, k=2 in pink, and k=3 in green.  The gray band is the best-fit modulation amplitude ($\pm 1\sigma$ error bars)  for 1.5-3.1 keVee, obtained using the log-likelihood method.  

The data support a modulation amplitude of $\sim 0.38 \pm 0.16$ cpd/kg/keVee in the high energy bin.  For the elastic scattering case considered here, only dark matter masses greater than $\sim 9$ GeV for k=1 and $\sim 10.5$ GeV for k=3 yield a modulation amplitude within a standard deviation of the measured value.   While a heavier dark matter mass increases the modulation amplitude at high energies, it also increases the unmodulated rate, making it conflict with the rates measured by CoGeNT.  For the wide range of halo and dark matter parameters considered here, only masses less than $\sim 7-9$ GeV are consistent with CoGeNT's unmodulated spectrum.  Unfortunately, none of these points give a sufficient modulation at high energies.  The results shown here emphasize the underlying tension in a dark matter interpretation of the CoGeNT data: namely, one must explain both an excess in the unmodulated rate below $\sim 0.9$ keVee and a significant modulation above $\sim 1.7$ keVee.  As Fig.~\ref{fig:nfwplots} highlights, a dark matter candidate scattering elastically off an equilibrated isotropic halo cannot satisfy both requirements.
However, this does not mean that it is impossible. In particular, a two-component halo that is described by a Maxwell-Boltzmann at low velocities and a stream at high velocities might allow consistency.  A modulation phase that is significantly different from 152 days would further point to a more exotic halo model.

\subsection{Model-Independent Comparisons}

In this subsection, we explore the constraints from other experiments on the CoGeNT modulation, assuming it arises from elastic dark matter.  Comparing rates between different direct detection experiments with different target nuclei is non-trivial because each probes a different range of dark matter velocities.  However, a means of comparing the results of different experiments independent of halo models has recently been proposed \cite{Fox:2010bz}.\footnote{For related work see \cite{Fox:2010bu}.}  For elastic spin-independent scattering, a signal in the range $[\elow^{(1)}, \ehigh^{(1)}]$ at Experiment 1 arises in Experiment 2 in the energy range
\be
[\elow^{(2)}, \ehigh^{(2)}] = \frac{\mu_2^2 M_T^{(1)}}{\mu_1^2 M_T^{(2)}}[\elow^{(1)}, \ehigh^{(1)}]~,
\label{eq:emap}
\ee
where $M_T^{(i)}$ is the mass of the target nucleus in each experiment and $\mu_i$ is the DM-nucleus reduced mass for each experiment. 
For a rate, $dR_1/dE_R$, observed at Experiment 1, the rate expected at Experiment 2 is
\be
\frac{dR_2}{dE_R}\left(E_2 \right) = 
\frac{C_T^{(2)}}{C_T^{(1)}} \frac{F_2^2(E_2)}{F_1^2\left(\frac{\mu_1^2\, M_T^{(2)}}{\mu_2^2 M_T^{(1)}} E_2\right)} 
\frac{dR_1}{dE_R}\left(\frac{\mu_1^2\, M_T^{(2)}} {\mu_2^2\, M_T^{(1)}}E_2\right)~.
\label{eq:ratequalSI}
\ee
Here,
\be
C_T^{(i)}=\kappa^{(i)}\left(f_p\, Z^{(i)}+f_n\, (A^{(i)}-Z^{(i)})\right)^2~,
\ee
where $\kappa$ is the mass fraction for the target element in question and $f_{p(n)}$ is the coupling strength of dark matter to protons (neutrons).  $F_i$ is the nuclear form factor for each experiment.

\begin{table}[t]
\begin{tabular}{|c|c||c|c|c|c|c|}
\hline
Bin & CoGeNT & Ge & Na (Q=0.3) & Si & O & Xe\\  \hline 
\multirow{2}{*}{1}&[0.5,0.9] 	&	[2.3,3.8]	&	[1.5,2.5]	&	[4.5,7.6]		& 	[5.8,9.9]	& 	[1.4,2.3] \\
& $0.90 \pm 0.72$ &   $0.23\pm0.18$	& $0.078\pm0.062$	& $0.035\pm0.028$	& $0.011\pm0.009$	& $0.72\pm0.58$\\ \hline
\multirow{2}{*}{2}&[0.9,1.5]	&	[3.8,6.1]		&	[2.5,4.0]	&	[7.6,11.9]		&	[9.9,15.6]	&	[2.3,3.7]\\
& $0.37 \pm 0.55$ &  $0.1\pm0.149$	& $0.035\pm0.052$	& $0.015\pm0.023$	& $0.005\pm0.008$	& $0.31\pm0.46$
 \\
 \hline
\multirow{2}{*}{3}&[1.5,2.3]	&	[6.1,8.9]	&	[4.0,5.8]	&	[11.9,17.5]	&	[15.6,22.8]	&	[3.7,5.4]\\
& $0.48 \pm 0.22$ & $0.136\pm0.063$	& $0.049\pm0.022$	& $0.021\pm0.01$	& $0.007\pm0.003$	& $0.41\pm0.19$
\\
 \hline
\multirow{2}{*}{4}&[2.3,3.1]	&	[8.9,11.6]	&	[5.8,7.6]	&	[17.5,22.8]	&	[22.8,29.8]	&	[5.4,7]\\
& $0.27 \pm 0.23$ & $0.08\pm0.068$	& $0.029\pm0.025$	& $0.013\pm0.011$	& $0.004\pm0.004$	& $0.23\pm0.2$
 \\
\hline
\end{tabular}
\caption{Predicted modulation amplitudes for example nuclear targets, given the best-fit values for CoGeNT assuming a Maxwellian phase. The units are in counts/day/kg/keVnr for all columns, except those labelled CoGeNT and Na where they are counts/day/kg/keVee; in the case of Na a quenching factor of Q=0.3 has been applied. The equivalent energy ranges and rates for other targets are shown, assuming $m_{\chi} = 7$ GeV and spin-independent scattering cross sections proportional to $A^2$. Note that we have not included detector efficiencies or mass fractions in any of the predicted rates.}
\label{tb:mapMW}
\end{table}

Tables~\ref{tb:mapMW} and \ref{tb:mapBF} show the ranges of energies at other experiments that correspond to the CoGeNT energy bins: [0.5, 0.9], [0.9, 1.5], [1.5, 2.3], and [2.3, 3.1] keVee.  Note that these energies are given in ``electron equivalent'' and correspond to [2.3, 3.8], [3.8, 6.1], [6.1, 8.9], and [8.9, 11.6] in nuclear recoil energies.  These tables also show how the CoGeNT modulation amplitude in each energy bin translates to other experiments, assuming a 7 GeV WIMP with spin-independent scattering proportional to $A^2$.  (Note that we have not included detector efficiencies or mass fractions in any of the predicted rates.)  Let us consider each experiment in turn.

\begin{figure}[tb] 
   \centering
   \includegraphics[width=3.5in]{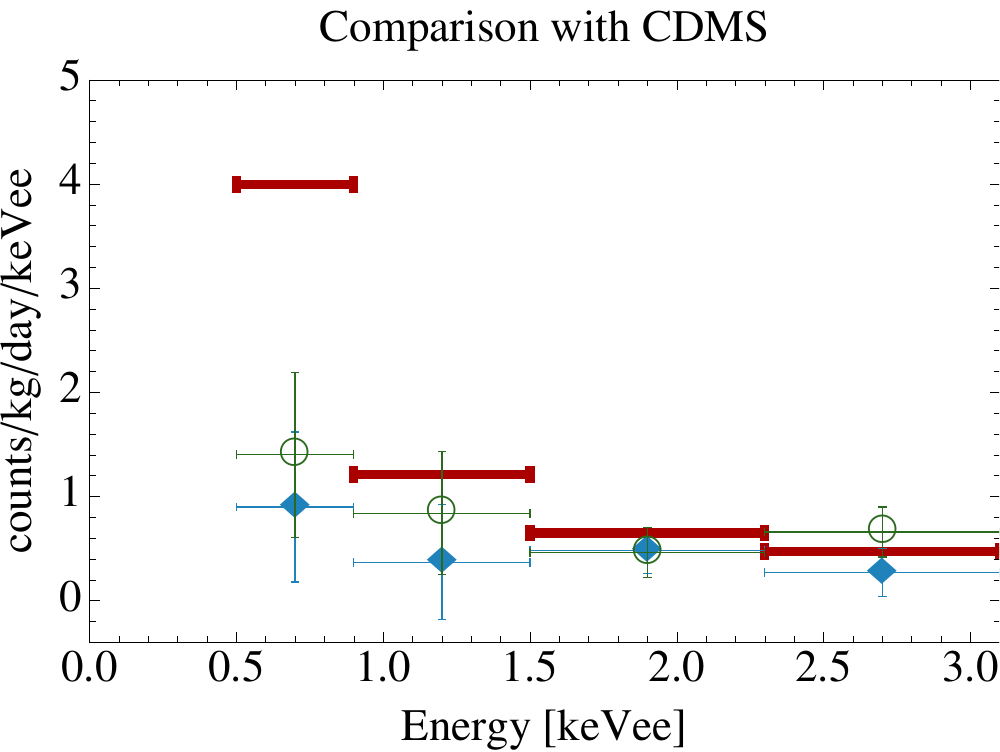}
   \caption{Upper limits from CDMS ({\it red}) compared with modulation rates from CoGeNT assuming a Maxwellian phase ({\it blue diamond}) and the overall best-fit phase ({\it green circle}).}
\label{fig:cdmscompare}
\end{figure}

\textbf{CDMS-Ge:} A direct comparison can be made between the CoGeNT and CDMS count rates because they both have germanium targets.  Using the results of the low-energy analysis of the CDMS experiment \cite{Ahmed:2010wy}, we calculate an upper limit for the rate in each detector such that it has a 1.3\% probability of having a lower rate.  This gives a probability of 10\% that any one of CDMS's eight detectors has a lower rate than is observed.  In each of the five energy bins, the strongest limit from all the detectors is chosen and we treat this as a 90\% confidence limit.\footnote{The probability that the particular detector that sets the limit has a strong downward fluctuation is small, and so the confidence is actually better than 90\%, but we treat it as a 90\% C.L. to be conservative.}  Figure~\ref{fig:cdmscompare} shows that the count rates at CDMS are not low enough to constrain the CoGeNT modulation.  However, the count rates {\em are} low enough that there should be modulation at a very high level in CDMS. Thus, even weak modulation constraints from CDMS could be very powerful; conversely, modulation should be apparent in a dedicated analysis, even with existing count rates.
\begin{table}[t]
\begin{tabular}{|c|c||c|c|c|c|c|}
\hline
Bin & CoGeNT & Ge & Na (Q=0.3) & Si & O & Xe\\  \hline 
\multirow{2}{*}{1}&[0.5,0.9] 	&	[2.3,3.8]	&	[1.5,2.5]	&	[4.5,7.6]		& 	[5.8,9.9]	& 	[1.4,2.3] \\
& $1.4 \pm 0.79$  & $0.36\pm0.2$ &	$0.12\pm0.07$ &	$0.054\pm0.03$ &	$0.018\pm0.01$ &	$1.1\pm0.6$ 
 \\ \hline
\multirow{2}{*}{2}&[0.9,1.5]	&	[3.8,6.1]		&	[2.5,4.0]	&	[7.6,11.9]		&	[9.9,15.6]	&	[2.3,3.7]\\
& $0.84 \pm 0.59$ & $0.23\pm0.16$ &	$0.079\pm0.055$ &	$0.035\pm0.024$ &	$0.012\pm0.008$ &	$0.70\pm0.49$ 
 \\
 \hline
\multirow{2}{*}{3}&[1.5,2.3]	&	[6.1,8.9]	&	[4.0,5.8]	&	[11.9,17.5]	&	[15.6,22.8]	&	[3.7,5.4]\\
& $0.46 \pm 0.24$ & $0.13\pm0.068$ &	$0.047\pm0.024$ &	$0.021\pm0.011$ &	$0.007\pm0.004$ &	$0.39\pm0.21$  \\
 \hline
\multirow{2}{*}{4}&[2.3,3.1]	&	[8.9,11.6]	&	[5.8,7.6]	&	[17.5,22.8]	&	[22.8,29.8]	&	[5.4,7]\\
& $0.66 \pm 0.24$ & $0.20\pm0.07$ &	$0.072\pm0.026$ &	$0.032\pm0.011$ &	$0.011\pm0.004$ &	$0.57\pm0.21$ \\
\hline
\end{tabular}
\caption{Same as Table~\ref{tb:mapMW}, except assuming a best-fit overall phase of 106 days.}
\label{tb:mapBF}
\end{table}

\textbf{CDMS-Si:} CoGeNT's modulation above 1.5 keVee should appear above threshold at CDMS-Si. Using the results from \cite{Filippini:2009zz}, with 88 kg-days and assuming an efficiency of 0.2, CoGeNT's observed modulation rates map into a minimum of $3.3 \pm 1.4$ ($5.0 \pm 1.5$) events for a Maxwellian (106 day) phase. CDMS-Si sees no events below 50 keVnr, so this predicted rate is borderline, but not excluded.  Two further conclusions can be made.  First, any signal should essentially be 100\% modulated (i.e., no sizeable constant piece). Second, invoking significant interference between proton and neutron couplings \cite{Chang:2010yk,Feng:2011vu} to reconcile CoGeNT with XENON is untenable. Taking $f_n=-0.7 f_p$ turns the relative boost between Si and Ge (the ratio of $C_T$'s) all the way up to 11, putting any observable modulation in the 1.5-3.1 range in serious conflict with the results form CDMS-Si.  Note, that the low energy calibration of the silicon detectors may be subject to corrections~\cite{Hooper:2010uy}, which would shift the energy threshold at CDMS-Si to higher recoil energies.  We have taken a fairly conservative approach in comparing CoGeNT to CDMS-Si, for instance only considering events above 11.9 keV in CDMS, and taking the efficiency to be a flat 20\%, and we do not expect the above tension to be greatly alleviated by potential threshold corrections.

\textbf{CRESST-O:}  CoGeNT's signal in the [0.5--1.5]~keVee region translates to an energy range that falls primarily below the threshold of many of CRESST's individual detectors. However, the [1.5-3.1]~keVee range should easily appear at CRESST above 15 keVnr in the oxygen band. With 600 kg-days of exposure and an efficiency of 80\%,  a modulated signal of $8.75 \pm 3.77$   ($13.1 \pm 4.0$) events is expected at CRESST, if the CoGeNT signal is due to a light elastic DM.  This is consistent with the $\sim 30$ events that have been observed at CRESST \cite{probsttalk}.  No significant modulation has yet been reported.

\textbf{XENON100:}  The energy range for CoGeNT's high-energy modulation is relevant for XENON100.  In particular, 1.5 keVee corresponds to 3.7~keVnr in XENON100 (again, for a 7 GeV WIMP), where the scintillation efficiency has been studied.  Taking an exposure of 1450 kg-days (48 kg $\times$ 100.9 days $\times$ overall efficiency of 0.3), we predict $1020 \pm 470$ ($970 \pm 510$) events from the third bin and $550 \pm 470$ ($1350 \pm 490$) events from the fourth bin {\em before taking into account the S1 cut}. 

Both of these bins are below the S1 threshold for XENON100 of four photo-electrons (PE), so we must calculate the probability that an upward fluctuation would occur. Both these bins occur above 3~keVnr, for which results are available for L$_{\text{eff}}$. Taking a value of L$_{\text{eff}}$=0.07 (approximately the lower boundary as measured by \cite{Plante:2011hw}), we predict efficiencies of 0.015 and 0.05 for the third and fourth bins, respectively.  With these efficiencies, one would have expected $15.4 \pm 7$ ($14.7 \pm 7.7$) and $27.4\pm 23.3$ ($66.9 \pm 24.3$) events based on CoGeNT's modulation in the third and fourth bins, respectively.  Reducing the Poisson efficiencies to an acceptable level requires that L$_{\text{eff}} \lsim$ 0.05 and $\lsim 0.04$ for the third and fourth bin, which lower the rates by a factor of approximately 3 and 6, respectively.  Whether such small values of L$_{\text{eff}}$ are possible or not is still a subject of active discussion in the literature~\cite{Collar:2010gd,XENON:2010er,Collar:2010gg,Aprile:2010um,Savage:2010tg,Collar:2010nx,Sorensen:2010hq,Manalaysay:2010mb,Collar:2010ht,Bezrukov:2010qa,Plante:2011hw,Collar:2011wq}

\begin{figure}[t] 
   \centering
   \includegraphics[width=3.2in]{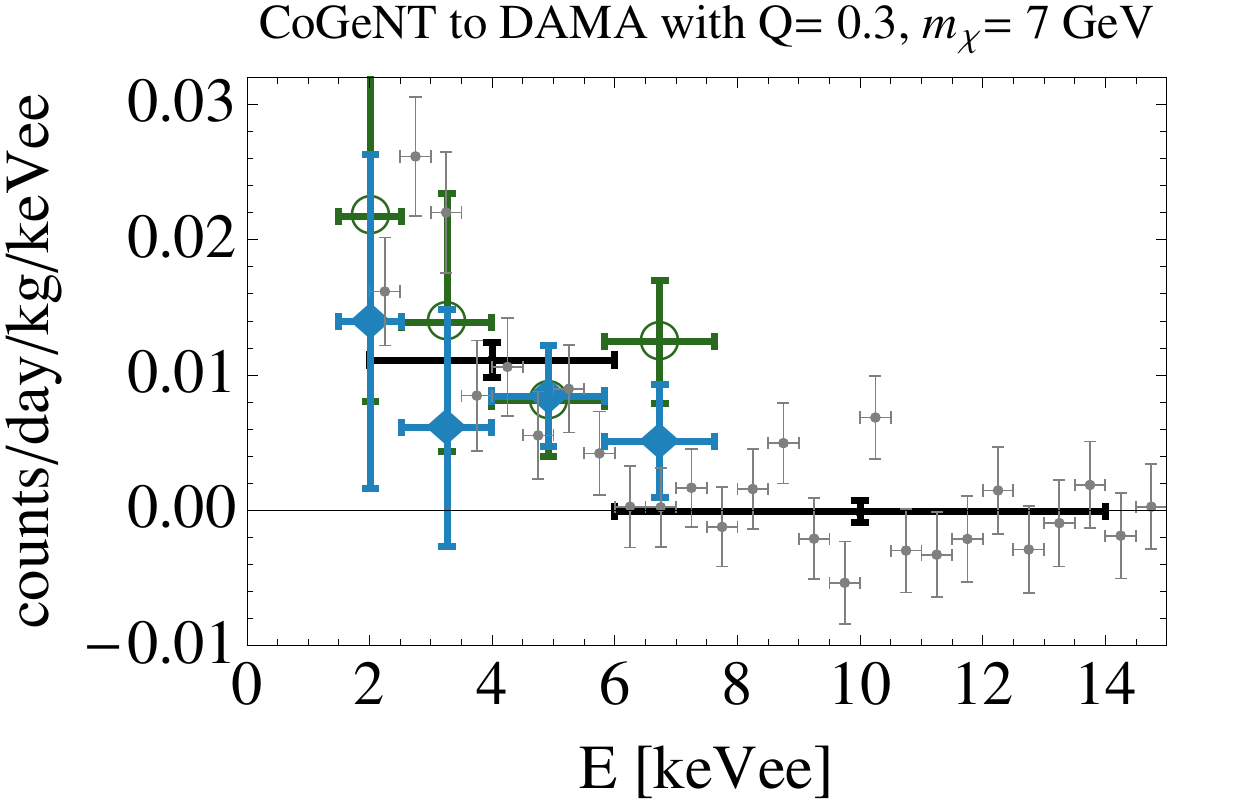} 
   \includegraphics[width=3.2in]{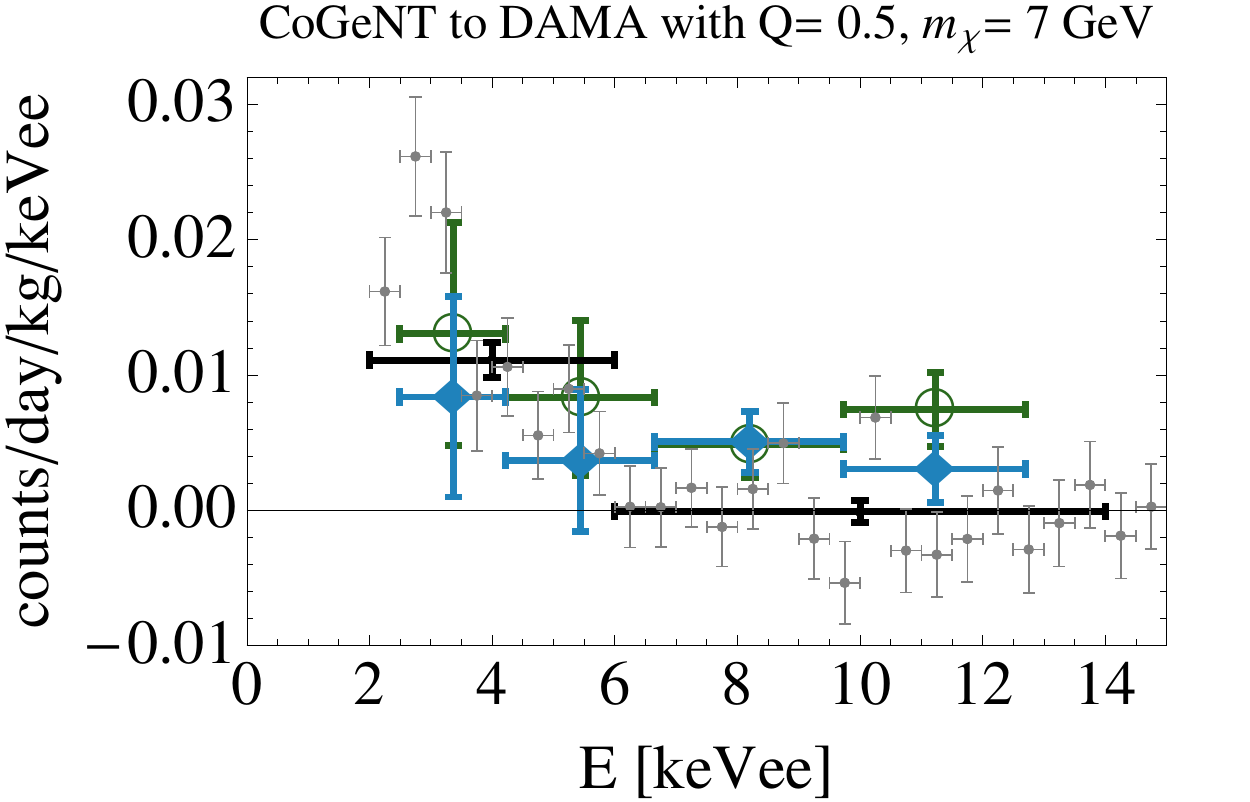} 
   \caption{Comparison of CoGeNT and DAMA modulation amplitudes.  The comparison is independent of astrophysics but assumes 7 GeV DM that scatters elastically and a quench factor in sodium of 0.3 (left plot) and 0.5 (right plot), since the energy bins are comparable to the DAMA resolution, we ignore the effects of energy smearing at DAMA.  In both plots, the blue diamonds (open green circles) denote the CoGeNT prediction for the modulation spectrum at DAMA, assuming a Maxwellian (best-fit) phase.  The black (gray) points are the results of a two bin (multi-bin) analysis by DAMA~\cite{Bernabei:2010mq}.}
   \label{fig:cogent2dama}
\end{figure}

\textbf{DAMA:}
Finally, we compare the measured modulation spectrum at DAMA with that of CoGeNT.   For a quenching factor of Q$_{\text{Na}}=0.3$ and $m_\chi \sim 7\gev$, the CoGeNT modulation energy range corresponds roughly with DAMA's (see Tables \ref{tb:mapMW},\ref{tb:mapBF}). The {\em total} modulation observed in the CoGeNT range 0.5-3.1 keVee yields a modulated rate of $0.04 \pm 0.017$ cpd/kg ($0.065 \pm 0.018$) for a Maxwellian (best-fit-106 day) phase, which compares with $0.0444 \pm 0.0052$ cpd/kg (for both MW and best-fit 146 day phases) at DAMA, assuming a conventional spin-independent mapping.  While the total modulated rate at CoGeNT is roughly consistent with DAMA's, the energy ranges at the two experiments do not line up exactly and the predicted rate at DAMA based on CoGeNT's signal is somewhat smaller than what is observed {\em independent of the astrophysical model}.  This is illustrated in Fig.~\ref{fig:cogent2dama}, since the energy bins are comparable in size to the energy smearing at DAMA we ignore the effect of smearing.  As previously noted~\cite{Chang:2010yk,Feng:2011vu}, taking the proton and neutron spin-independent couplings to interfere can favor light targets and correct this. Such effects can happen through interactions via heavy fermionic mediators \cite{Feng:2011vu} or through $Z'$s \cite{Fox:2011qd}. However, taking $f_n = -0.7 f_p$ to maximize the suppression at XENON boosts the modulation at DAMA relative to CoGeNT by a factor of six relative to the case of $f_n = f_p$, which seems in conflict with the data in hand.

We also consider the specific scenario of \cite{Hooper:2010uy}, with a 7 GeV WIMP and a large quench factor in sodium, Q$_{\text{Na}}$ = 0.5 \cite{Hooper:2010uy}. In this case, the lowest two bins at CoGeNT map roughly into the DAMA range 2-6 keVee, yielding $0.019\pm 0.015$ ($0.036\pm 0.016$) cpd/kg for a Maxwellian (106 day) phase.  The higher energy bins (1.5-3.1 keVee) map into the range of 6.6-12.7 keVee at DAMA, and predict a modulation of $0.02 \pm 0.0085$ ($0.03\pm 0.009$) cpd/kg to be compared with the observed $-0.0008 \pm 0.0064$ cpd/kg observed in the 6-14 keVee range, see Fig.~\ref{fig:cogent2dama}. Consequently, this mass and quenching factor are in tension with the high energy data.  While this conflict is at the 2 (2.7) $\sigma$ level for Maxwellian (106 day) phase, ignoring it requires one to essentially ignore modulation that is as significant as the modulation that one is taking seriously. As a result, we believe that this particular scenario does not give a good fit to the data.

\subsubsection{Summary of Halo-Independent Comparisons}
To summarize, this subsection compares results between different experiments in a manner that is independent of astrophysical uncertainties.  The most direct comparison is between CoGeNT and CDMS-Ge; the results of the two experiments are consistent only if CDMS's rate is modulated at nearly 100\%.  Such a modulation should be easily visible in the CDMS data.

 Ultimately, while there is rough agreement between the size of the CoGeNT modulation and the DAMA modulation, the energy range over which the modulation is spread is in conflict with previous interpretations \cite{Hooper:2010uy} invoking a large sodium quenching factor, because this disregards the modulation at high energies, which is as statistically significant as that in the lower energy range.
 
The presence of modulation in the high energy range results in the greatest tensions with other experiments. The absence of a signal at CDMS-Si requires the signal to be highly modulated, while XENON100 should have seen a signal unless L$_{\text{eff}}$ is significantly smaller than the measurements of \cite{Plante:2011hw}.  Invoking interference between protons and neutrons to alleviate XENON100 constraints exacerbates tensions with CDMS-Si.

The comparisons made in this subsection are only valid in the context of spin-independent scattering where the rate is proportional to 
$A^2$.  We have seen that any interpretation in this context is challenging because of the high-energy modulation in CoGeNT.  Other models, such as spin-dependent or inelastic interactions, fall outside the scope of this analysis.  

\section{Conclusions}

We have studied the event rate from the CoGeNT experiment over the first 458 days of running.  CoGeNT continues to see an unexplained excess at low energies and now claims evidence for a modulating spectrum.  The presence of modulation makes the possibility of a WIMP explanation more compelling, but also requires a serious and thorough discussion of its implications for theory and other direct detection experiments.

In this work, we confirm the original findings of \cite{Aalseth:2011wp} of a modulation in the low-energy (0.5-3.1 keVee) data of CoGeNT, with a significance of  99\% or 2.6$\sigma$ (99.7\% or 3$\sigma$) for a Maxwell-Boltzmann (best-fit) phase.  However, the details of the modulation make it somewhat confusing. The significance for modulation in the low energy range (0.5-1.5 keVee), where the signal from a light WIMP would be present, is only 87\% (90\%) for a Maxwell-Boltzmann (best-fit) phase. In contrast, the significance is 97.7\% (98.3\%)  in the higher energy range (1.5-3.1 keVee).  The absence of significant modulation below $\sim 1.5$ keVee is not troubling on its own, because modulation fractions can be larger at higher energies, and cosmogenic backgrounds at 0.9-1.5 keVee are significant. Nonetheless, one cannot claim that the modulation in the low-energy regime gives evidence for a model while simultaneously disregarding the much more significant modulation in the high-energy regime $-$ i.e., if a model does not explain a significant part of the high energy modulation, it cannot be claimed to explain the modulation. The significance of modulation in the 1.5-3.1 keVee region is confirmed by a variety of statistical techniques, and is an important contributor to the total modulation significance in the 0.5-3.1 keVee energy range.

Many elastic dark matter models with Maxwellian halos do not give any significant modulation above 1.5 keVee for masses below $\sim 7-9$ GeV. We have performed a broad study of elastic WIMP scenarios with Maxwellian and NFW-consistent halos and have found no models that give modulation at high energies, while not exceeding the {\em unmodulated} count rate at lower energies.  Halo models with debris flows \cite{Lisanti:2011as} or streams might allow for this, however.  Our best-fit WIMP point with a Maxwell-Boltzmann halo does not significantly improve the fit compared to a constant background.  As a result, we believe that attempts to understand the CoGeNT modulation with a Maxwellian halo are likely to be unreliable and lead to erroneous conclusions.

To that end, we have attempted to employ techniques that are independent of the halo model \cite{Fox:2010bz} when comparing the CoGeNT results to other experiments.  We find that a direct comparison to CDMS-Ge allows the modulation, but predicts that a significant modulation should appear at CDMS as well. Note that small errors in energy, while important for interpreting a rapidly falling background, should not affect a general modulation analysis such as this one.

Other experiments also provide insight on the modulation in the 1.5-3.1 keVee range.  In particular, the absence of a signal at CDMS-Si suggests that any signal in this range should be $\sim$ 100\% modulated.  A study of XENON100 is intriguing, because the modulation appears in a range that has been more directly calibrated with measurements of L$_{\text{eff}}$. We find that any significant modulation in the 1.5-3.1 range should have shown up at the 10-30+ event level at XENON100, unless the value of L$_{\text{eff}}$ is significantly lower than what has been found by \cite{Plante:2011hw}.

A halo model-independent comparison to DAMA shows generally {\em good} agreement between the event rates for a light WIMP and Q$_{\text{Na}} \approx 0.3$.  However, the modulation at CoGeNT is in conflict with DAMA if one assumes Q$_{\text{Na}} = 0.5$, disfavoring previous interpretations utilizing a Maxwellian halo.

Inelastic dark matter models can provide a good fit to the modulation, but not the exponentially falling rate at low energies. iDM in the presence of stream(s) or debris flows could lead to the highly-modulated signal that is observed. If the signal is only present in April, it could explain why XENON100 did not observe a significant rate, as data taking in that period was limited by excess noise.

In summary, the modulation at CoGeNT is an intriguing piece of the puzzle, but raises as many questions through the high-energy modulation as it answers.  Future data from CoGeNT is imperative for understanding the presence or absence of modulation at low and high energies and thus, shedding light on whether--and what type of--dark matter could give rise to it.

\noindent {\bf Note Added}
At the time of the completion of the manuscript, the authors became aware of related papers~\cite{Schwetz:2011xm, Farina:2011pw}, which explore implications of the CoGeNT data for particle and astrophysical models.  The authors of \cite{Schwetz:2011xm} carried out an analysis on the two bin data presented in \cite{Aalseth:2011wp}, while \cite{Farina:2011pw} use the full data set analysed in this work.  Our conclusions are in qualitative agreement.

After this manuscript had been submitted to the e-print arXiv, the CoGeNT collaboration released new preliminary results which indicate that part of the low-energy excess events may be due to surface backgrounds~\cite{Collar-TAUP}. If confirmed, this would imply that the parameter region favored by the CoGeNT event spectrum would become larger and move to lower cross sections~\cite{Collar-TAUP, Kopp:2011yr, Kelso:2011gd}.

\vskip 0.2in
\noindent {\bf Acknowledgements}
We are especially grateful to Juan Collar for making the CoGeNT data publicly available and for his support in explaining many aspects of it during our analysis.
We also wish to thank Rouven Essig, Roni Harnik, Graham Kribs, Rafael Lang, Josh Ruderman, Tracy Slatyer, Natalia Toro, Michael Witherell and Itay Yavin for useful discussions.  ML acknowledges support from the Simons Postdoctoral Fellowship and the LHC Theory Initiative. NW is supported by NSF grant \#0947827, as well as support from the Amborse Monell Foundation. This research was supported in part by the National Science Foundation under Grant No. NSF PHY05-51164. Fermilab is operated by Fermi Research Alliance, LLC, under Contract DE-AC02-07CH11359 with the United States Department of Energy.


\begin{appendix}
\section{CoGeNT data and background}
\label{sec:gapsandbacks}

\begin{table}[b]
    \begin{tabular}{ | c | c | c | c | c | c | c | c | }
      \hline
    & & \multicolumn{3}{c|}{\vphantom{\huge X}L-shell}		& \multicolumn{3}{c|}{ K-shell}  	 \\ 	 		
    \hline
    \vphantom{\huge X}  Isotope 	& $t_{1/2}$ (days)		& $N	_{\text{atoms}}$	& $E_0$ (keVee) 	& $\sigma$ (keVee)& $N	_{\text{atoms}}$	& $E_0$ (keVee) 	& $\sigma$ (keVee)	\\ 
     \hline
     \hline
     ${}^{73}$As 	&80			& 14.7 			& 1.41 			& 0.0777		&  133		& 11.1	& 0.120	\\ \hline
     ${}^{68}$Ge	&271		& 736			& 1.30			& 0.0770		&  6460		& 10.4	&  0.117	\\ \hline
     ${}^{68}$Ga	&271		& 60.9			& 1.19			& 0.0764		&  553		& 9.66	& 0.114	\\ \hline
     ${}^{65}$Zn	& 244		& 243			& 1.10			& 0.0759		& 2250		& 8.98	& 0.112	\\ \hline
     ${}^{56}$Ni	& 5.9			& 1.78			& 0.926			& 0.0749		& 17.2		& 7.71	& 0.107	\\ \hline
     ${}^{56,58}$Co& 71		& 10.9			& 0.846			& 0.0744		& 107		& 7.11	& 0.104	\\ \hline
     ${}^{57}$Co	& 271		& 2.98			& 0.846			& 0.0744		& 29.3		& 7.11	& 0.104	 \\ \hline
     ${}^{55}$Fe	& 996		& 51.8			& 0.769			& 0.0740		& 489		& 6.54	& 0.102	\\ \hline
     ${}^{54}$Mn	& 312		& 24.3			& 0.695			& 0.0736		& 238		& 5.99	& 0.100	\\ \hline
     ${}^{51}$Cr	& 28			& 3.38			& 0.628			& 0.0732		& 33.5		& 5.46	& 0.0975	\\ \hline
     ${}^{49}V$	& 330		& 17.2			& 0.564			& 0.0728		&172		& 4.97	& 0.0953	\\ \hline
    \end{tabular}
   \caption{Data used to model the cosmogenic background for both L-shell and K-shell decays.  $N_{\text{atoms}}$ is the number of atoms in the detector expected to decay via each mode (before efficiency corrections), $E_0$ is the binding energy, $\sigma$ is the energy resolution, and $t_{1/2}$ is the half-life for the decay.}
    \label{tab: CosmoTable}
\end{table}

The known background in the energy region of interest arises from cosmogenic electron capture events, which can be modeled as a sum of decaying gaussians.    A radioactive element, with Gaussian peak at $E_0$, width $\sigma$, and half-life $t_{1/2}$ has an energy spectrum of
\begin{equation}
f_{\text{peaks}}(E, t)= N_{\text{atoms}}\frac{ \log 2}{t_{1/2}} 2^{-t/t_{1/2}}\frac{1}{\sqrt{2\pi} \sigma}e^{-(E-E_0)/2\sigma^2}~.
\label{eq:cosmopeaks}
\end{equation}
Here, $N_{\text{atoms}}$ is the total number of atoms expected to decay in the detector.  The number of K-shell decays is obtained by fitting to the peaks in the high energy spectrum.  These numbers are provided in the public release of the CoGeNT data and are summarized in Table~\ref{tab: CosmoTable}, where they have been corrected for efficiencies.  We have been able to reproduce $N_{\text{atoms}}$ for the ${}^{68}$Ge peak at 10.4 keVee using our own fitting procedure.  The binding energies for the K-shell peaks are given in~\cite{Aalseth:2010vx} and the resolution is given by the formula in~\cite{Aalseth:2008rx}, with parameters from~\cite{Aalseth:2010vx}. 

The expected number of L-shell decays at lower energies is related to the number of K-shell decays~\cite{Bahcall:1963zza}; therefore, the third column in Table~\ref{tab: CosmoTable} can be obtained from the measured quantities in the sixth column.  The L-shell cosmogenics are relevant from $\sim$ 0.5-1.7 keVee, where the dark matter signal is expected to dominate.  The dominant contribution in the first year of running comes from ${}^{68}$Ge, followed by ${}^{65}$Zn.  

The observed energy spectrum at CoGeNT is affected by an efficiency that captures the decreasing sensitivity to signal near the experiment's energy threshold.  The efficiency is approximately flat at $\sim 0.87$ between 0.7 - 3.0 keVee, and then drops to 0.75 by 0.5 keVee.  Above 4 keVee, it is 0.94.  In this work, a spline interpolation of the efficiency data points is used, denoted by $f_{\text{eff}}(E)$.     

The CoGeNT data run spanned 458 days, of which 442 were live.  The time gaps in which no data were taken must be properly accounted for in a study of annual modulation.  To account for these outages, the following function is introduced:
\begin{equation}
f_{\text{gaps}}(t)=\left(\Theta(67-t)+\Theta(t-74)\right)\left(\Theta(101-t)+\Theta(t-107)\right)\left(\Theta(305-t)+\Theta(t-308)\right)~,
\end{equation}
where $t$ is measured in days.  Therefore, the complete expression for the spectrum of cosmogenic peaks is 
\begin{equation}
f_{\text{cosmo}} (E, t) = f_{\text{peaks}}(E, t) f_{\text{gaps}}(t) f_{\text{eff}}(E).
\label{eq:totalcosmo}
\end{equation}

\end{appendix}
\newpage
\bibliography{cogentanalysis}

\end{document}